\documentclass[twocolumn]{aastex631}

\usepackage{float}
\usepackage{graphicx}
\usepackage{subfigure}
\usepackage{placeins}
\usepackage{amsmath}
\usepackage{booktabs}        
\usepackage{threeparttable}

\received{October 13, 2025}
\revised{November 7, 2025}
\accepted{November 19, 2025, to Publications of the Astronomical Society of the Pacific}

\begin{document}

\title{Machine learning classification of baseband data of CHIME FRBs}

\correspondingauthor{Tetsuya Hashimoto}
\email{tetsuya@phys.nchu.edu.tw}

\author[0009-0005-9739-5540]{Mohanraj Madheshwaran}
\affiliation{Department of Physics, National Chung Hsing University, South District, 402, Taichung, Taiwan}

\author[0000-0001-7228-1428,sname='Hashimoto']{Tetsuya Hashimoto}
\affiliation{Department of Physics, National Chung Hsing University, South District, 402, Taichung, Taiwan}

\author[0000-0002-6821-8669]{Tomotsugu Goto}
\affiliation{Institute of Astronomy, National Tsing Hua University, Kuang-Fu Road,  30013, Hsinchu, Taiwan}

\author[0000-0002-7300-2213]{William J. Pearson}
\affiliation{National Centre for Nuclear Research, Pasteura 7, 02-093 Warszawa, Poland}

\author[0000-0003-4860-0737]{Murthadza Aznam}
\affiliation{Department of Physics, Faculty of Science, Universiti Malaya, Kuala Lumpur 50603, Malaysia}

\author[0000-0002-8560-3497,sname='Ho']{Simon C.-C. Ho}
\affiliation{Research School of Astronomy and Astrophysics, The Australian National University, Canberra, ACT 2611, Australia}

\author[0009-0001-9195-7494,sname='Vavillakula Venkataramana Rao']{Vignesh V.V. Rao}
\affiliation{Department of Physics, National Chung Hsing University, South District, 402, Taichung, Taiwan}

\author[0000-0003-3747-9847,sname='Gajendran']{Sridhar Gajendran}
\affiliation{Institute of Astronomy, National Tsing Hua University, Kuang-Fu Road,  30013, Hsinchu, Taiwan}
\affiliation{National Centre for Radio Astrophysics (NCRA–TIFR), Pune, 411007, India}





\begin{abstract}
Fast Radio Bursts (FRBs) are bright millisecond radio pulses. 
Their origin is still unknown in the field of astronomy. 
A notable distinction among FRBs is that some sources repeat, while others appear to be non-repeating events. 
Interestingly, repeating FRBs tend to exhibit broader temporal widths and narrower spectral bandwidths compared to non-repeat events, 
suggesting they may arise from different physical mechanisms. 
However, current radio telescopes have limited coverage and sensitivity, which hinders a complete survey with continuous long-term monitoring. 
This issue makes it difficult to confirm repeat activity and potentially leads to misclassification of repeaters as non-repeaters; these are referred to as repeater candidates. 
To address this, machine learning techniques have emerged as a useful tool for classifying distinct FRB types in previous studies.  
In this study, we utilize the CHIME/FRB baseband catalog with three orders of magnitude better time resolution than the intensity catalog.
Measured fluences are available in the baseband catalog, while only upper limits are reported in the intensity catalog.  
We apply machine learning to the baseband catalog to evaluate classification outcomes. 
We identify 15 repeater candidates among 122 non-repeating FRBs in the baseband catalog. 
Additionally, our classification identifies 31 sources previously categorized as repeater candidates as non-repeaters, highlighting a significant 
difference from the prior work. 
Of these repeater candidates, 14 overlap with previous findings, while 1 is newly identified in this work. 
Notably, one of our candidates was confirmed as a repeater by CHIME/FRB. 
Follow-up observations for the 14 candidates are highly encouraged. 
\end{abstract}



\section{Introduction} \label{sec:intro}
Researchers have proposed many theoretical models in recent years to explain fast radio bursts (FRBs) \citep{2019PhR...821....1P}. 
\cite{2007Sci...318..777L} defines FRBs as the millisecond-duration astronomical transients that cause bright pulses during radio observations. 
Up to this point, there are more than 800 FRB events that have been detected by observations \citep{2023Univ....9..330X}.
Most of them occur at extragalactic distances \citep[e.g.,][]{2013Sci...341...53T}. 
Among them, approximately 120 FRBs have been identified with known host galaxies (localized) \citep[e.g.,][]{2017Natur.541...58C, 2020TNSAN.160....1P, 2019Sci...366..231P, 2019Natur.572..352R, 2020Natur.581..391M, 2020Natur.577..190M}. 
On the other hand, FRB 200428 is the only FRB known to be associated with a magnetar in our Galaxy \citep[e.g.,][]{2020Natur.587...54C, 2020Natur.587...59B}.

{Researchers classify FRBs into two types: repeaters and non-repeaters
\citep[e.g.,][]{2019Natur.572..352R, 2020MNRAS.498.3927H}. On an observational basis, any FRB source detected emitting multiple bursts is categorized as a repeater, while a source with no such multiple detections is considered a non-repeater. 
The repeater FRB is often discussed as being associated with a magnetar \citep{2020Natur.587...59B}, and the non-repeater FRB is often discussed as being associated with cataclysmic events \citep{2019NatAs...3..928R}.
Repeating FRBs can be localized precisely, enabling us to identify their host galaxies and environments. For instance, a repeating FRB (FRB 121102) was localized to its host galaxy \citep{2024Ap&SS.369...59L}. Approximately 60 FRB sources are known to repeat, and the vast events are non-repeater \citep[e.g.,][]{2023ApJ...947...83C, 2019ApJ...885L..24C, 2019Natur.566..235C, 2019ApJ...887L..30K, 2020ApJ...891L...6F, 2022Natur.602..585K, 2022Natur.606..873N, 2022Natur.609..685X}.  
In comparison to non-repeater FRBs, repeaters exhibit statistically significant differences, including longer durations, narrower bandwidths, and a more complex burst structure, often consisting of multiple sub-bursts \citep{2021ApJ...923....1P}.  Therefore, proper classification between repeaters and non-repeaters is important because they might originate from different progenitors.

However, misclassification could happen due to the observational limitations. 
Repeating FRBs could be missed (i) if they happen outside observational time windows 
or (ii) if the fluence of FRBs is below the telescope's sensitivity.
Therefore, an FRB identified as non-repeating could actually be a repeating FRB, but the repetitions are missed in the observation due to the limitations mentioned above. This issue hampers the correct understanding of FRB origins. 
Moreover, such limitations could cause misclassification between non-repeaters and repeaters.
Hence, a source not detected repeating may still be an actual repeater. In other words, it is challenging to ensure that non-repeater samples are entirely free from contamination by repeaters.

\textcolor{red}{\cite{2019NatAs...3..928R}} investigates the volumetric occurrence rate of nearby non-repeating FRBs to find that the FRB volumetric rate exceeds the rates of candidate cataclysmic progenitor events, including core-collapse supernovae, neutron-star mergers, magnetars, etc. 
They conclude that most FRBs, including apparent non-repeater, originate from repeaters, based on the rate of volumetric occurrence. 
A consistent conclusion is reported by \cite{2024MNRAS.52711158Y} by using the time evolution of the FRB detection rates. Yet, the observations show a larger number of non-repeater events than repeater events.

Proper classification of FRBs requires extensive observation, e.g., long-term monitoring with wide field-of-view telescopes. 
However, it is difficult in practice. 
Researchers have been trying to classify these two types of FRBs for decades. 
For example, \cite{2020MNRAS.494.2886H} use only two parameters of repeaters and non-repeaters to present their different distributions.
They utilize rest-frame intrinsic duration and time-integrated luminosity to find different data distributions between repeaters and non-repeaters. 
As mentioned above, repeaters exhibit longer durations and narrower bandwidths \citep{2021ApJ...923....1P}. 
However, \citet{2020MNRAS.494.2886H} did not utilize the bandwidth information for their classification. 
Therefore, including more parameters could give a more reliable classification.

Machine learning can effectively handle as many parameters as are available.
It may facilitate the classification of FRBs without long-term monitoring and minimal human intervention. 
For instance, machine learning was applied by \cite{2023MNRAS.518.1629L} for the classification of repeaters and non-repeaters. 
Their model classifies most repeater FRBs correctly, attributing the differences to distinct underlying mechanisms, without long-term observation and minimal human intrusion.
In addition, several studies have been conducted to identify repeater candidates through machine-learning approaches. 
For instance, deep neural networks were used by \cite{agarwal2020searches} to classify the repeater candidates in the observed data from the Australian Square Kilometre Array Pathfinder (ASKAP).

The Canadian Hydrogen Intensity Mapping Experiment/Fast Radio Burst (CHIME/FRB) catalog 1 \citep[also known as the intensity catalog;][]{2021ApJS..257...59C} is currently the largest and homogeneous FRB sample detected with a single instrument. This dataset was obtained from a single observation under uniform selection effects \citep{2021ApJS..257...59C}. This catalog contains 536 FRBs. 
This marked the first huge dataset, which includes both repeaters and non-repeaters. Therefore, the CHIME/FRB catalog 1 would be suitable for machine-learning analyses. Moreover, \cite{2025arXiv250906208K} employed a deep learning approach using the latest CHIME/FRB Catalog 2 to classify repeaters and non-repeaters.  

\cite{2022MNRAS.509.1227C} have identified 188 repeater candidates from 
the CHIME/FRB catalog 1
by using unsupervised machine learning. 
The CHIME/FRB catalog 1 is referred to as the intensity catalog in this paper.
\cite{2023MNRAS.522.4342Y} applied an unsupervised machine learning technique to both a parameter-based catalog and image data of the CHIME/FRB intensity catalog.
They aimed to identify repeater candidates and investigate the relationship between the results of the parameter-based catalog and image data.
On the other hand, 
\cite{2024ApJ...969..145C} enhanced this existing intensity catalog by providing baseband measurements for 140 of these FRBs. 
Further details of the baseband catalog can be found in \cite{2024ApJ...969..145C}. 
This baseband catalog comprises 12 repeater bursts and 128 non-repeater bursts.


In this work, we use unsupervised machine learning on this baseband catalog to identify repeater candidates. The misclassification problem could be resolved with long-term monitoring of each FRB source with high sensitivity. However, such observations are too expensive. Therefore, an alternative approach is important to resolve the misclassification issue. This work aims to identify repeater candidates using the baseband catalog and compare our results with those in \cite{2022MNRAS.509.1227C}. Once proven, the ML classification would be extremely useful because it does not require expensive long-term monitoring.

The structure of this paper is as follows. Section \ref{sec:param-collect} introduces the baseband data and selected parameters, while \ref{sample} details the sample selection. Section \ref{machine} details the machine learning model, hyperparameter optimization, optimized model configuration, and model evaluation. Section \ref{result} reports the unsupervised machine learning results and the identification of repeater candidates. Section \ref{discussion} provides a discussion of the astrophysical implications. Our conclusions are presented in Section \ref{conclusion}.

\section{Parameter selection and data collection}\label{sec:param-collect}
\subsection{The data of Baseband catalog}\label{sec:advantage}
In this work, we used the CHIME/FRB baseband catalog and intensity catalog for machine learning classification. The baseband catalog is an enhanced version of the intensity catalog with improved measurements of FRBs. Further details of the baseband catalog can be found in \cite{2024ApJ...969..145C}. 
In the intensity catalog, the flux and fluence are calibrated from the dynamic spectrum \citep{2023AJ....166..138A}.
These two parameters are lower limits in the intensity catalog \citep{2024ApJ...969..145C}. 
In contrast, in the baseband catalog, these values are measured from total intensity data (burst intensity recorded during observation) stored in single-beam files \citep{2024ApJ...969..145C}.
The baseband catalog also has more precise measurements of celestial coordinates, observed dispersion measure (DM), and higher time resolution than those in the intensity catalog.
The observed scattering time scale ranges from 30 $\mu$\text{s} to 13 \text{ms} at 600 MHz \citep[e.g.,][]{2025ApJ...979..160S}, highlighting the importance of the high time resolution. 
Overall, the baseband data have improved time resolution and fluence measurements.

\subsection{Parameter selection}\label{sec:selection}
In this research, we aim to incorporate as many relevant parameters as possible to enrich the sensitivity and robustness of the results. 
In total 16 parameters, which are relevant to FRB properties, are publicly available in the intensity catalog and baseband catalog \citep{2021ApJS..257...59C, 2024ApJ...969..145C}.
We chose 11 parameters out of 16, which are included in observational and model-dependent parameters, namely: (1) spectral index, (2) spectral running, (3) highest frequency, (4) lowest frequency, (5) peak frequency, (6) flux, (7) fluence, (8) boxcar width, (9) scattering time, (10) redshift, and (11) radio energy.
We calculated the redshift and radio energy using astronomical models, which are called model-dependent parameters in this work. Other parameters are observed parameters, recorded during radio observations of FRBs. The spectral index, spectral running, highest frequency, lowest frequency, and peak frequency are taken from the intensity catalog \citep{2024ApJ...969..145C}. 
Flux and fluence are attained from the baseband catalog \citep{2024ApJ...969..145C}. 
Boxcar width and scattering time are obtained from the baseband-data morphology \citep{2025ApJ...979..160S}.
Seven FRBs, namely FRB 20181220A, FRB 20181228B, FRB 20190202B, FRB 20190517C, FRB 20190612A, FRB 20190626A, and FRB 20190628C, do not have flux and fluence values in the baseband catalog. 
Therefore, the values of flux and fluence for these seven FRBs are employed from the intensity catalog. 

In the morphology study of baseband data \citep{2025ApJ...979..160S}, the duration and scattering time are not available for FRB 20190612A, FRB 20190628C, and FRB 20190627D. Hence, the duration and scattering time values for these three FRBs are acquired from the intensity catalog.


\cite{2022MNRAS.509.1227C} did machine learning classification by using the intensity catalog. On the other hand, we used the baseband catalog in this work. As the baseband catalog includes updated measurements of 140 FRBs, the FRB samples in our dataset are also present in their catalog. This circumstance presents an opportunity to identify common repeater candidates, and therefore, to compare our results with \cite{2022MNRAS.509.1227C} because we have adopted a similar machine learning approach.
They included similar time domain parameters, such as the width of sub-bursts and the rest-frame intrinsic duration, both of which are practically identical to the boxcar width.
Also, the boxcar width is measured from the baseband catalog with a better time resolution, whereas the width of sub-bursts and the rest-frame intrinsic duration are measured from the intensity catalog. 
Therefore, we excluded those two parameters used in \cite{2022MNRAS.509.1227C} from our analysis due to their similarities and the poor time resolution in the intensity catalog. 
Following \cite{2025ApJ...980..185S}, we include both flux and fluence in our analysis because flux is sensitive to the instant brightness of FRBs and fluence is an estimate of integrated brightness in a given time duration.

\subsubsection{The observational parameters}\label{observational}
The observational parameters adopted in our analysis are summarized in the following list.

\textbf{1. Spectral Index:} 
It represents the spectral shape of each burst. Precisely, spectral index shows the relationship between the flux and frequency of the FRBs \citep{2019ApJ...872L..19M}. \cite{2024ApJS..271...49F} developed an effective model for spectral energy distribution using physical and heuristic parameters of
CHIME data that contains pulsars and FRBs. The spectral index used in this work is derived from their model. 

\textbf{2. Spectral Running:} This parameter represents an additional term to describe a non-power-law shape of an FRB spectrum, including a Gaussian-like function and asymmetric peaks on either end of the band \citep{2021ApJ...923....1P}.

\textbf{3. Highest Frequency (MHz):} 
This is the maximum value of the frequency range measured by using the channelized baseband data \citep{2024ApJ...969..145C}.

\textbf{4. Lowest Frequency (MHz):} 
This is the minimum value of the frequency range measured by using the channelized baseband data \citep{2024ApJ...969..145C}.

\textbf{5. Peak Frequency (MHz):} This parameter represents the peak of an FRB spectrum in the frequency domain. The channelized baseband data was used to estimate this parameter \citep{2024ApJ...969..145C}.

\textbf{6. Flux (Jy):} The flux indicates the peak in the band-averaged light curve of an FRB. 
Flux is measured by using total intensity data stored in the single-beam files generated during the final stage of the automated pipeline \citep{2024ApJ...969..145C}. 

\textbf{7. Fluence (Jy$\cdot$ms):} 
Fluence refers to the time-integrated flux over the duration of an FRB. It was measured by using total intensity data stored in the single-beam files generated during the final stage of the automated pipeline \citep{2024ApJ...969..145C}. 


\textbf{8. Boxcar Width (s):} The boxcar width manifests the total duration of an FRB.
This measurement includes the effects of instrumental broadening, scattering, and redshift, and remains consistent across each FRB event \citep{2021ApJS..257...59C}.

\textbf{9. Scattering Time (s):} This parameter represents the pulse broadening time due to scattering at 600 MHz with the redshift broadening effect retained \citep{2021ApJS..257...59C}.

\subsubsection{The model-dependent parameters}\label{model-dependent}
The redshift and radio energy are model-dependent parameters. 
The measurement of observed dispersion measure ($\mathrm{DM}_{\mathrm{obs}}$) describes the electron density integrated over the physical distance $ds$ \citep[e.g.,][]{2003ApJ...598L..79I,2004MNRAS.348..999I,2020Natur.581..391M}.
The dispersion measure of intergalactic medium  ($\mathrm{DM}_\mathrm{{IGM}}$) is one of the components of $\mathrm{DM}_{\mathrm{obs}}$, and it is expected to have a strong dependence on redshift \citep[e.g.,][]{2014PhRvD..89j7303Z}. 
The radio energy of the FRB is indicated by the integration of its observed fluence over frequency \citep[e.g.,][]{2022MNRAS.511.1961H}. 
In this work, redshift and radio energy were calculated using these models. 
Therefore, they are considered model-dependent parameters.

Spectroscopic redshifts (spec-$z$) were used directly for nine FRBs with available measurements:  FRB 20181223C, FRB 20190418A, and FRB 20190425A \citep{2024ApJ...971L..51B}; FRB 20181225A, FRB 20181226A, FRB 20190605A, and FRB 20190605B \citep{2020Natur.577..190M}; FRB 20190611A, FRB 20190626A \citep{2023ApJ...950..134M}.
The redshift of the rest of the samples is estimated by using their dispersion measures and equatorial coordinates obtained from the baseband data. 
For the calculation of redshift and radio energy, we followed the same method mentioned in \cite{2019MNRAS.488.1908H} and \cite{2022MNRAS.511.1961H}, respectively. 
The brief description of the calculation method for redshift and radio energy is provided below.

\textbf{10. Redshift:} This parameter gives information about the source distance. It was estimated based on their observed dispersion measure $\mathrm{DM}_{\mathrm{obs}}$. The observed dispersion measure is composed of multiple contributions. It is described as follows:
\begin{equation}\label{dm_equation}
\mathrm{DM}_\mathrm{{obs}} = \mathrm{DM}_\mathrm{{MW}}(b,l) + \mathrm{DM}_\mathrm{{halo}} + \mathrm{DM}_\mathrm{{IGM}}(z) + \mathrm{DM}_\mathrm{{host}}(z),
\end{equation}
where $\mathrm{DM}_\mathrm{{MW}}(b,l)$ is the DM contribution of Milky Way.
$\mathrm{DM}_\mathrm{{halo}}$ is the DM contribution of the Galactic halo.
$\mathrm{DM}_\mathrm{{IGM}}(z)$ is the DM contribution of extragalactic plasma \citep[e.g.,][]{2020Natur.581..391M}.
$\mathrm{DM}_\mathrm{{host}}(z)$ is the DM contribution of a host galaxy.

\citet{2024ApJ...969..145C} has revised the celestial coordinates in the baseband catalog. 
We convert these updated coordinates to Galactic coordinates using the \texttt{astropy.coordinates} module \citep{2013A&A...558A..33A}. 
We apply the YMW16 \citep{2017ApJ...835...29Y} electron-density model to calculate the $\mathrm{DM}_\mathrm{{MW}}$ by integrating along the line of sight up to 25 kpc. 
We use $\mathrm{DM}_\mathrm{{halo}}=65 \: \text{pc}\: \text{cm}^{-1}$, following average values reported in previous studies \citep[e.g.,][]{2019MNRAS.485..648P}. 
Following literature \citep{2018Natur.562..386S}, we assume 
\begin{equation}
\mathrm{DM}_\mathrm{{host}} = \frac{50.0}{(1+z)} \: \text{pc} \: \text{cm}^{-3}.
\end{equation}
For an FRB at a more distant Universe, its signal passes through more ionized material in space. 
Therefore, $\mathrm{DM}_\mathrm{{IGM}}$ can be used as an estimate of each FRB's redshift. 
The cosmic average of $\mathrm{DM}_\mathrm{{IGM}}$ can be calculated using an analytical formula that depends on redshift, along with certain cosmological parameters \citep{2014PhRvD..89j7303Z} as follows.

\begin{equation}
\begin{aligned}
\mathrm{DM}_{\mathrm{IGM}}(z) =\ & \Omega_b\: \frac{3H_0c}{8\pi G m_p} \times \\
&\hspace{-7em} \int_{0}^{z} 
\frac{(1+z')f_{\mathrm{IGM}}(z')\left(Y_H X_{e,H}(z') + \frac{1}{2} Y_p X_{e,He}(z')\right)}%
{\left[\Omega_m (1+z')^3 + \Omega_{\Lambda} (1+z')^{3[1+\omega(z')]}\right]^{1/2}}\, dz',
\end{aligned}
\end{equation}
where, $X_{e, H} \: ~\text{and}\: X_{e, He} $ represent the ionization fractions of hydrogen and helium, respectively. 
We adopt their mass fractions of $Y_{H}=\frac{3}{4}\: \text{and}\: Y_{p}=\frac{1}{4}$, respectively. The equation of state describing dark energy is given by $\omega$.
We assume $\omega=-1$, which corresponds to no-redshift evolution of the equation of state of dark energy \citep{2001IJMPD..10..213C, 2003PhRvL..90i1301L}. 
The IGM is assumed to be fully ionized for a reasonable redshift range up to $z$ $\sim$ 3, hence $X_{e, H}=1\: \text{and}\: X_{e, He}=1$. 
We note that redshifts of our samples are all below $z=3$ (see section \ref{sample} for the details).
In accordance with previous work \citep{2014PhRvD..89j7303Z}, we incorporated $f_{IGM}=0.9\: ~\text{at}\: z>1.5\: ~\text{and}\: ~f_{IGM}=0.053z+0.82\: \text{at}\: z\leq 1.5$. 
By combining the expression for $\mathrm{DM}_\mathrm{{IGM}}\: \text{and}\: \mathrm{DM}_\mathrm{{host}}$, Equation \eqref{dm_equation} becomes a function of redshift. 
Solving this function for a given $\mathrm{DM}_\mathrm{{obs}}$ yields an estimate of the redshift for each FRB.

The method outlined above is described in detail in \cite{2020MNRAS.498.3927H}. In this work, we follow the same approach for estimating redshifts from observed dispersion measures. Readers are encouraged to read the reference for a comprehensive explanation of the underlying assumptions and derivations.

\textbf{11. Radio Energy (\textit{erg}):}
This parameter represents the rest frame isotropic radio energy.
It was calculated from the observed fluence. The brightness of FRBs is indicated by the integration of fluence over frequency. As a first step, the observed energy ($E_{obs}$) for each FRB is calculated by integrating the fluence over frequency. It is expressed as follows:

\begin{equation}
E_{obs} = \text{fluence} \times \left(\frac{400 \times 10^6}{{\rm Hz}}\right).
\end{equation}

We employ a fixed 400 MHz frequency width in the rest frame to provide a fair comparison of measured energy across various redshifts. The following expression provides the relevant frequency difference $\Delta \nu_{obs, itg}$ in the observer frame :

\begin{equation}
\Delta \nu_{obs, itg} = \frac{400}{(1+z)} {\rm MHz} 
\end{equation}

The observed energy integration is defined as follows:

\scalebox{0.8}{$ 
\displaystyle
E_{\text{obs},400} = 
\begin{cases}
F_\nu \left( \dfrac{4 \times 10^8}{\text{Hz}} \right) & (\Delta \nu_{obs, \text{itg}} \geq \Delta \nu_{obs, \text{FRB}}) \\[1ex]
F_\nu \left( \dfrac{4 \times 10^8}{\text{Hz}} \right)
\left( \dfrac{\Delta \nu_{obs,\text{itg}}}{\Delta \nu_{obs, \text{FRB}}} \right) & (\Delta \nu_{obs, \text{itg}} < \Delta \nu_{obs,\text{FRB}})
\end{cases}
$}

\begin{itemize}
    \item $F_\nu$ is the observed fluence from the baseband catalog.
    \item $\Delta \nu_{\text{obs, FRB}}$ is the observed bandwidth of the FRB, calculated as:\\
    $\Delta \nu_{\text{obs, FRB}} = \text{Highest frequency} - \text{Lowest frequency}$
    \item $\frac{\Delta \nu_{\text{obs, itg}}}{\Delta \nu_{\text{obs, FRB}}}$ represents the approximate energy that has overflowed out of the rest-frame.
\end{itemize}

Next, we calculate the rest-frame radio energy $(E_{rest, 400})$ for each FRB. It is expressed as:

\begin{equation}
    E_{rest, 400} = 4 \pi d_l^2 \left(\frac{E_{obs, 400}}{1+z}\right),
\end{equation}
where, 
$d_l$ represents the luminosity distance. 
The luminosity distance was calculated for each FRB using its corresponding redshift. 
The above method on the methodology for computing the radio energy as detailed by \cite{2022MNRAS.511.1961H}. We use the same methodology in this work. Readers are recommended to go to the original reference for a thorough explanation of the process and its underlying assumptions.

We applied a log$_{10}$ transformation to all parameters except spectral index and spectral running. These two parameters can take negative values in our dataset on a log scale because they represent the spectral indices of FRBs' spectra. Hence, we did not apply the log$_{10}$ transformation for these two parameters. 
Depending on the adopted ranges of physical parameters, the actual values change significantly, which might affect the clustering results \citep[e.g.,][]{2023MNRAS.522.4342Y}. Therefore, to remove this possible effect, we applied z-score standardization for all of the input parameters before training our model. This process converts each data point to show how many standard deviations it is away from the mean.
\section{Sample selection}\label{sample}
Our preliminary dataset consists of 140 baseband FRBs before applying any selection criteria. 
The redshift calculation method ($z_{\rm baseband}$) is explained in the section \ref{model-dependent}. 
We also calculate the redshift of FRBs using DM observations and Galactic coordinates provided in the intensity catalog ($z_{\rm intensity}$), following the same method described in section \ref{model-dependent}.
Because some FRB coordinates changed in the baseband catalog, their $\mathrm{DM}_\mathrm{{MW}}$ changed \footnote{The typical positional accuracy of the CHIME/FRB intensity catalog is $\sim$
15$'$–30$'$. Due to the interferometric nature of CHIME, the point-source localization can be improved by mapping the signal intensity around the initial FRB detection. One can fit a model of the expected telescope response to the intensity map to obtain a more accurate position in the baseband catalog \citep{2021ApJ...910..147M}. Due to the improvement of the positional accuracy, some FRBs' coordinates changed in the baseband catalog.}.
Consequently, $z_{\rm baseband}$ can be different from $z_{\rm intensity}$. 
we plot the redshift difference ($\Delta z = z_{\rm intensity} - z_{\rm baseband}$) between intensity and baseband catalogs against (1 + $z_{\rm baseband}$). We compare the redshift differences between intensity and baseband catalogs using the function of (1 + $z_{\rm baseband}$). We showed this difference in Fig. \ref{fig:redshift_difference}, and it was discovered that three FRB samples, FRB 20190419B, FRB 20190607B, and FRB 20190624B, deviated from the equality line. This deviation indicates that these three FRB samples have more variations in redshift between the baseband and intensity catalogs.
These three outliers could affect the structure of the baseband dataset in the high-dimensional space, including the relationship between the redshift, spectral shape, and the other FRB parameters.  
Therefore, to reduce the impact of these outliers and ensure the robustness of the machine learning model, we exclude these three samples from our analysis. 


\FloatBarrier
\begin{figure}[tbp]
    \centering
    \includegraphics[width=0.47\textwidth]{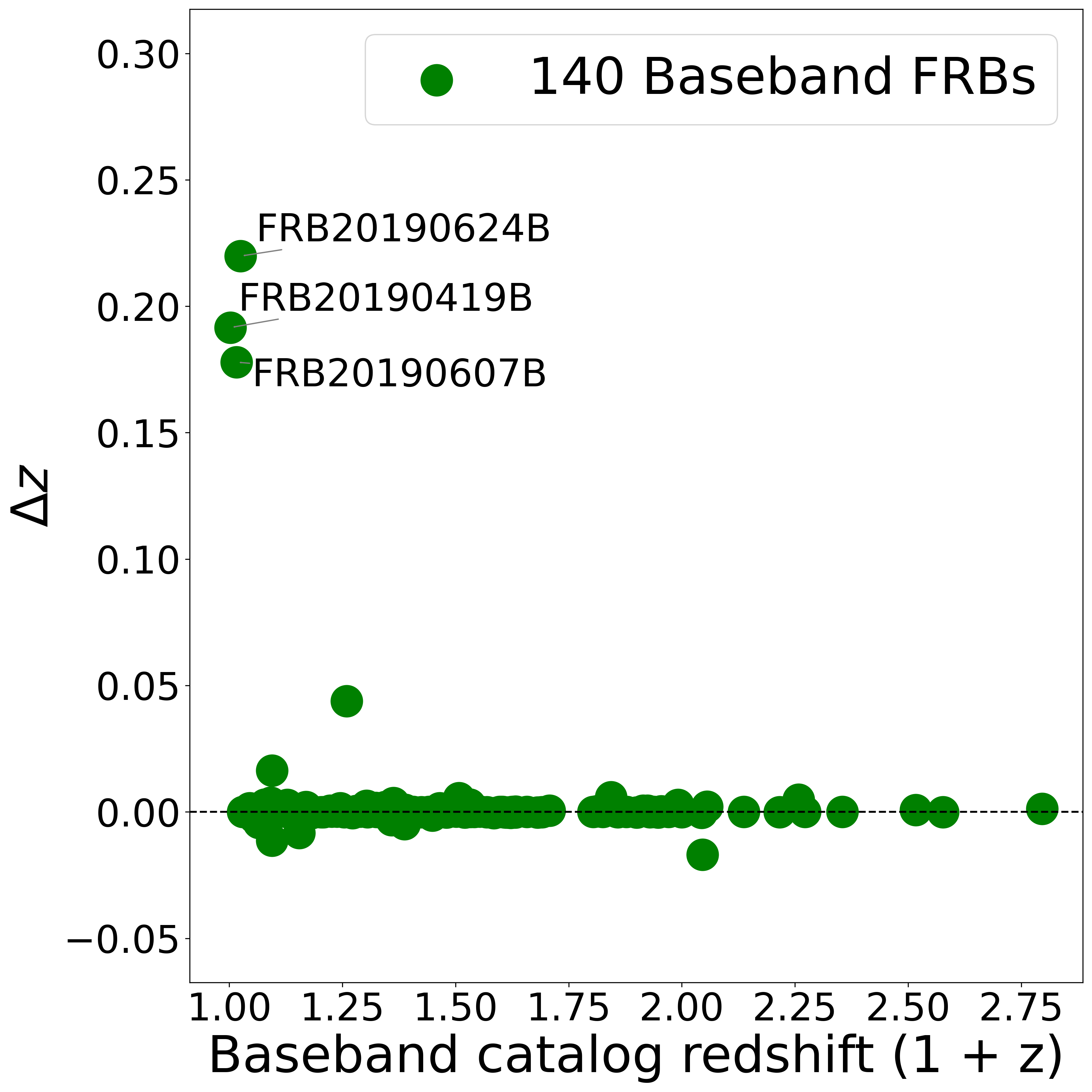}
    \caption{
    The redshift difference between intensity and baseband catalogs, plotted against the baseband redshift as $(1 + z_{\rm baseband})$.
    The difference is calculated by $\Delta z = z_{\rm intensity} - z_{\rm baseband}$. 
    Highlighted FRBs exhibit significant deviations from the line of equality (horizontal dashed line), suggesting inconsistencies in their redshifts and potentially low positional accuracy.
    }
    \label{fig:redshift_difference}
\end{figure}

Furthermore, three non-repeaters: FRB 20181220A, FRB 20190517C, FRB 20190613B, and one repeater: FRB 20190625E, have negative redshift values in our calculation. 
At the low-$z$ universe, the expected $\mathrm{DM}_{{IGM}}(z)$ is smaller than at the high-$z$ universe. $\mathrm{DM}_{{IGM}}(z)$ is derived by subtracting $\mathrm{DM}_\mathrm{{MW}}$ and $\mathrm{DM}_\mathrm{{host}}$ from $\mathrm{DM}_\mathrm{{obs}}$. Therefore, the uncertainties of $\mathrm{DM}_\mathrm{{MW}}$ and $\mathrm{DM}_\mathrm{{host}}$ affect $\mathrm{DM}_\mathrm{{IGM}}$ at lower redshifts more significantly than at higher redshifts. The DM-derived redshift can be negative within this uncertainty at the low-$z$ universe. 
Therefore, we decided to exclude them from our analysis. 
Overall, seven FRBs were excluded from further analysis, resulting in a final sample set that contains 11 repeaters and 122 non-repeaters, for a total of 133 FRB bursts.
We note that we adopt the measurements of the first sub-burst of each FRB in this work. 
The first sub-burst represents the first-arrived sub-burst for each FRB event \citep{2021ApJS..257...59C}.
The distributions of the 11 parameters for repeaters and non-repeaters are shown in Figure \ref{fig:histogram}. 


\begin{figure*}[tbp]
    \centering
    \includegraphics[width=0.3\textwidth]{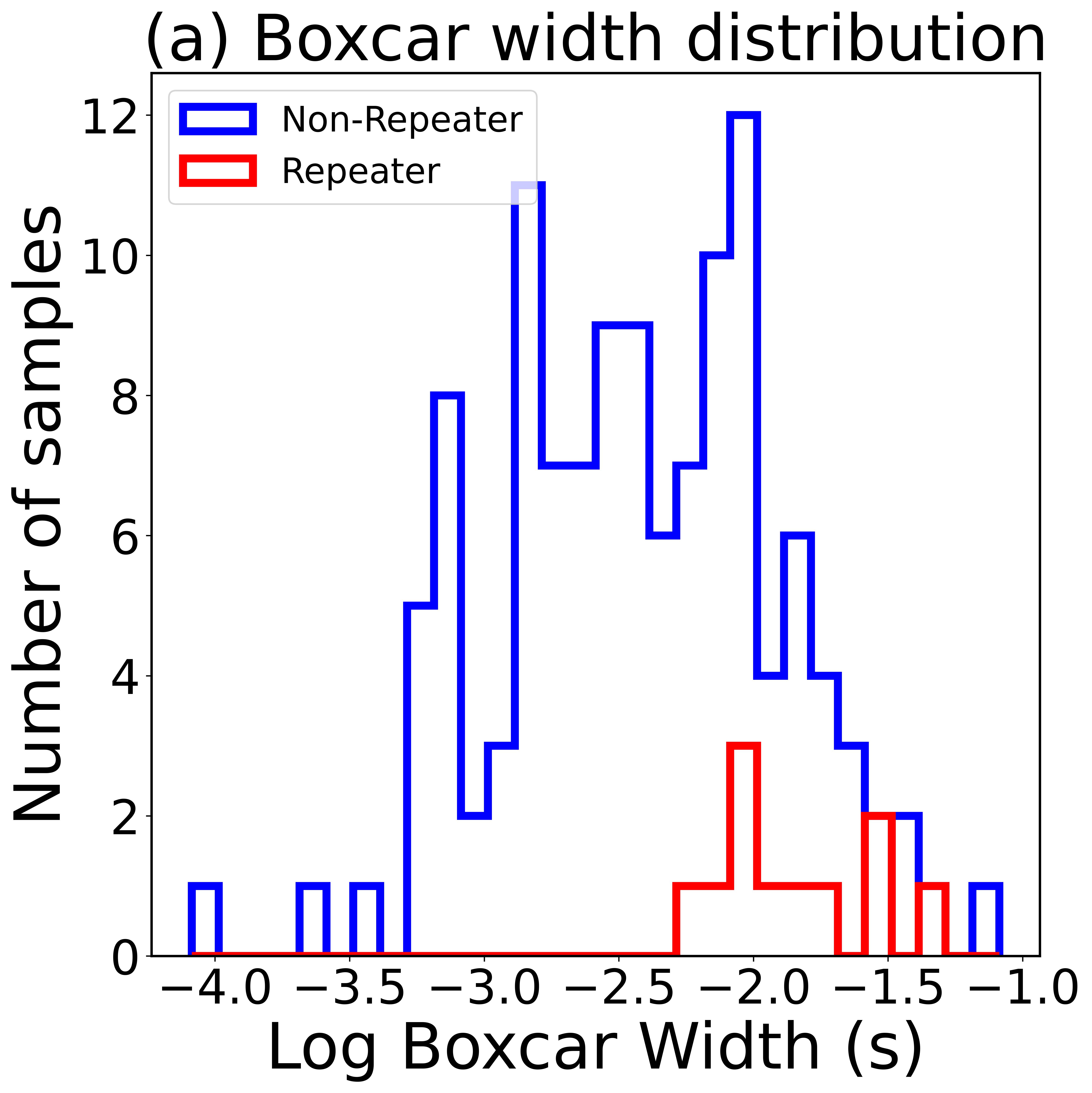}
    \includegraphics[width=0.3\textwidth]{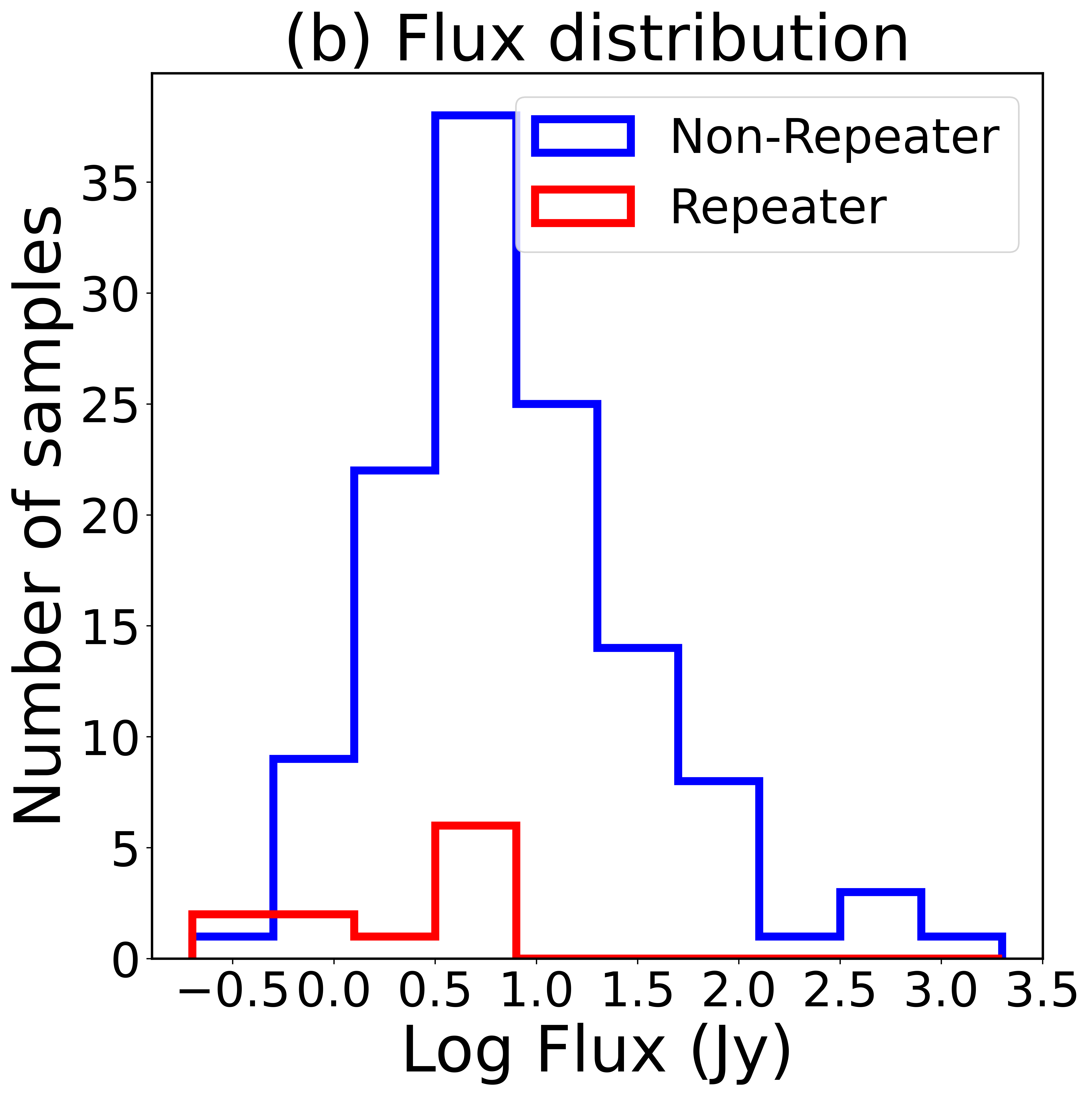}
    \includegraphics[width=0.3\textwidth]{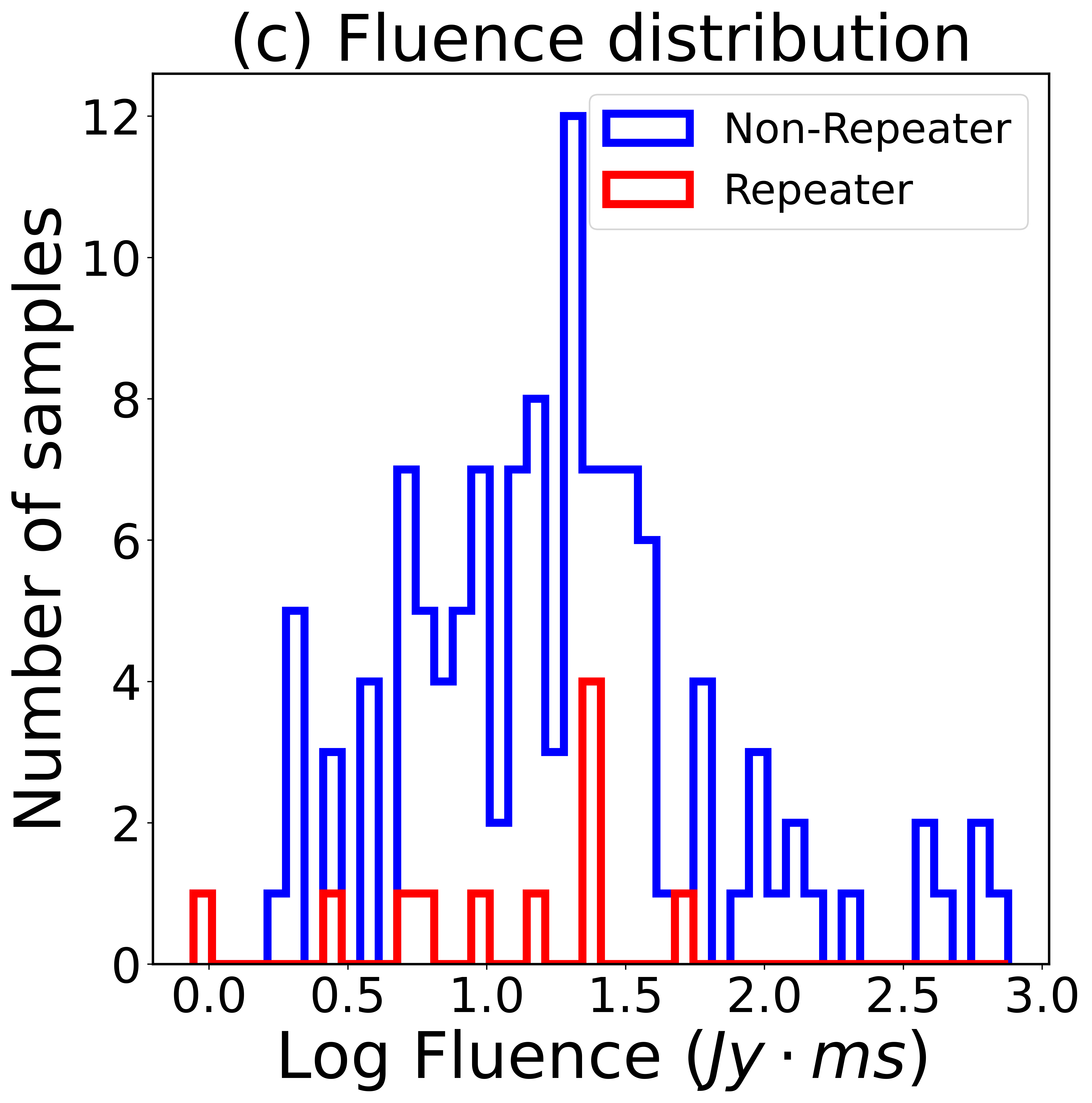}

    \includegraphics[width=0.3\textwidth]{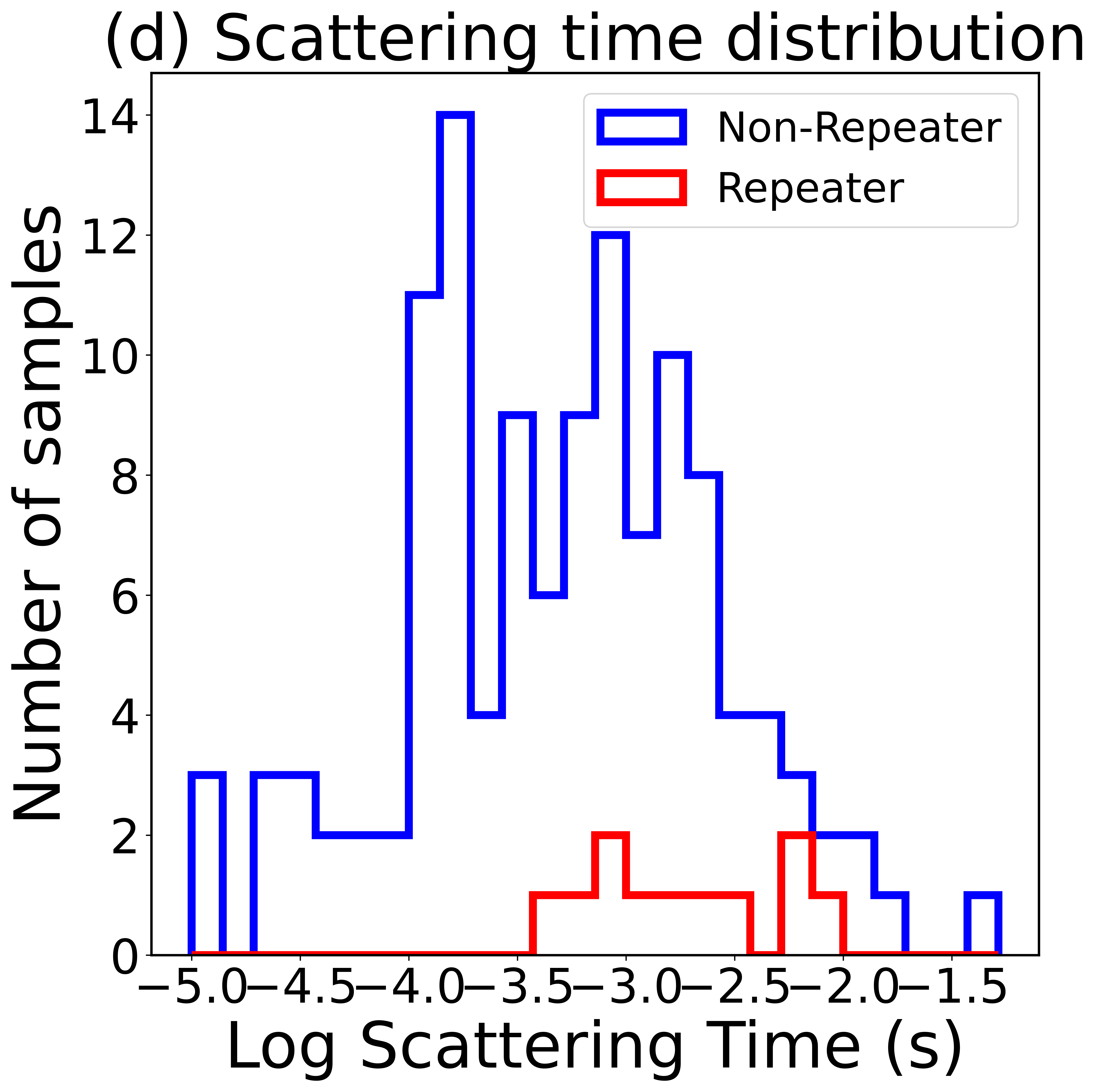}
    \includegraphics[width=0.3\textwidth]{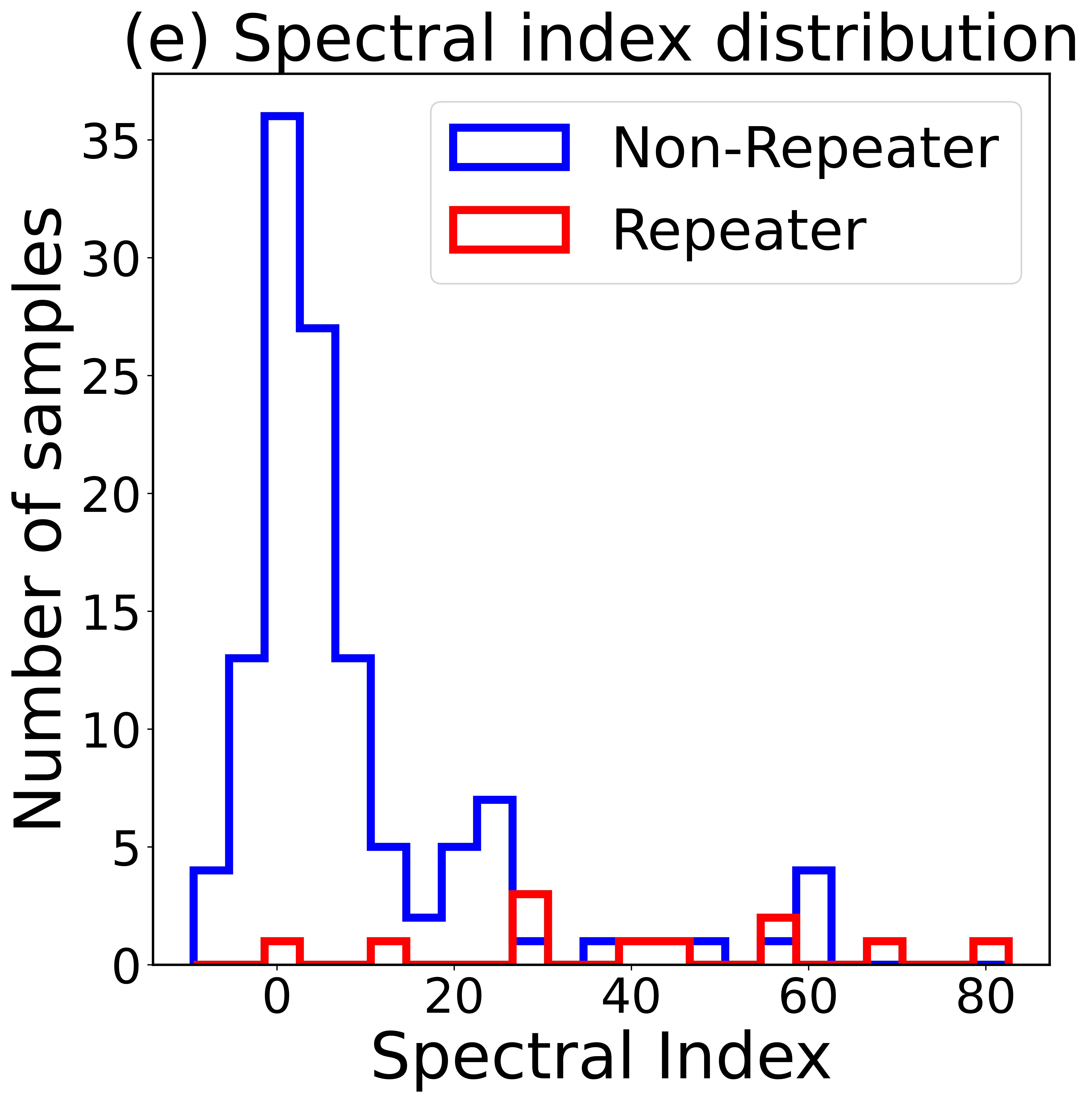}
    \includegraphics[width=0.3\textwidth]{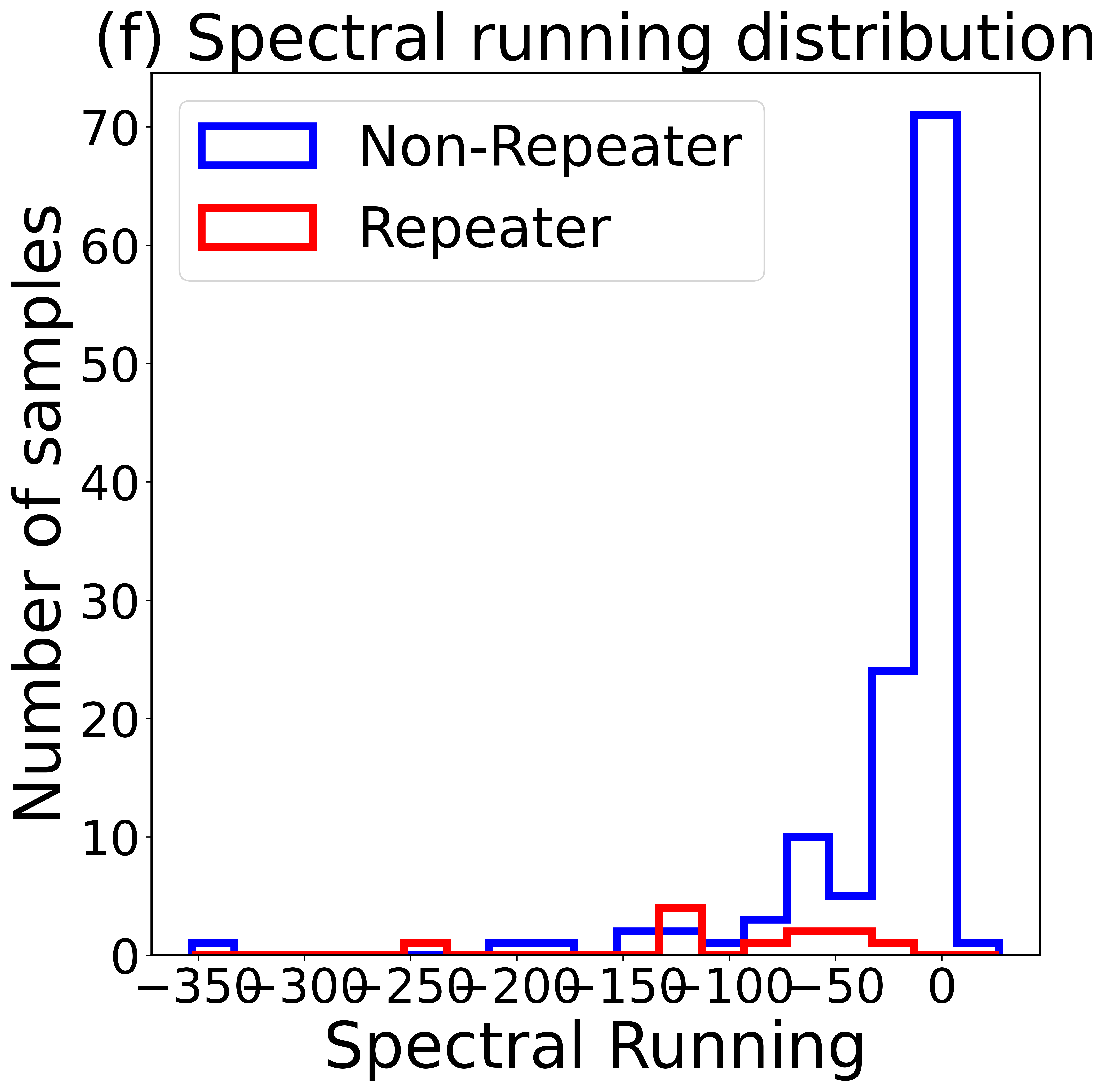}

    \includegraphics[width=0.3\textwidth]{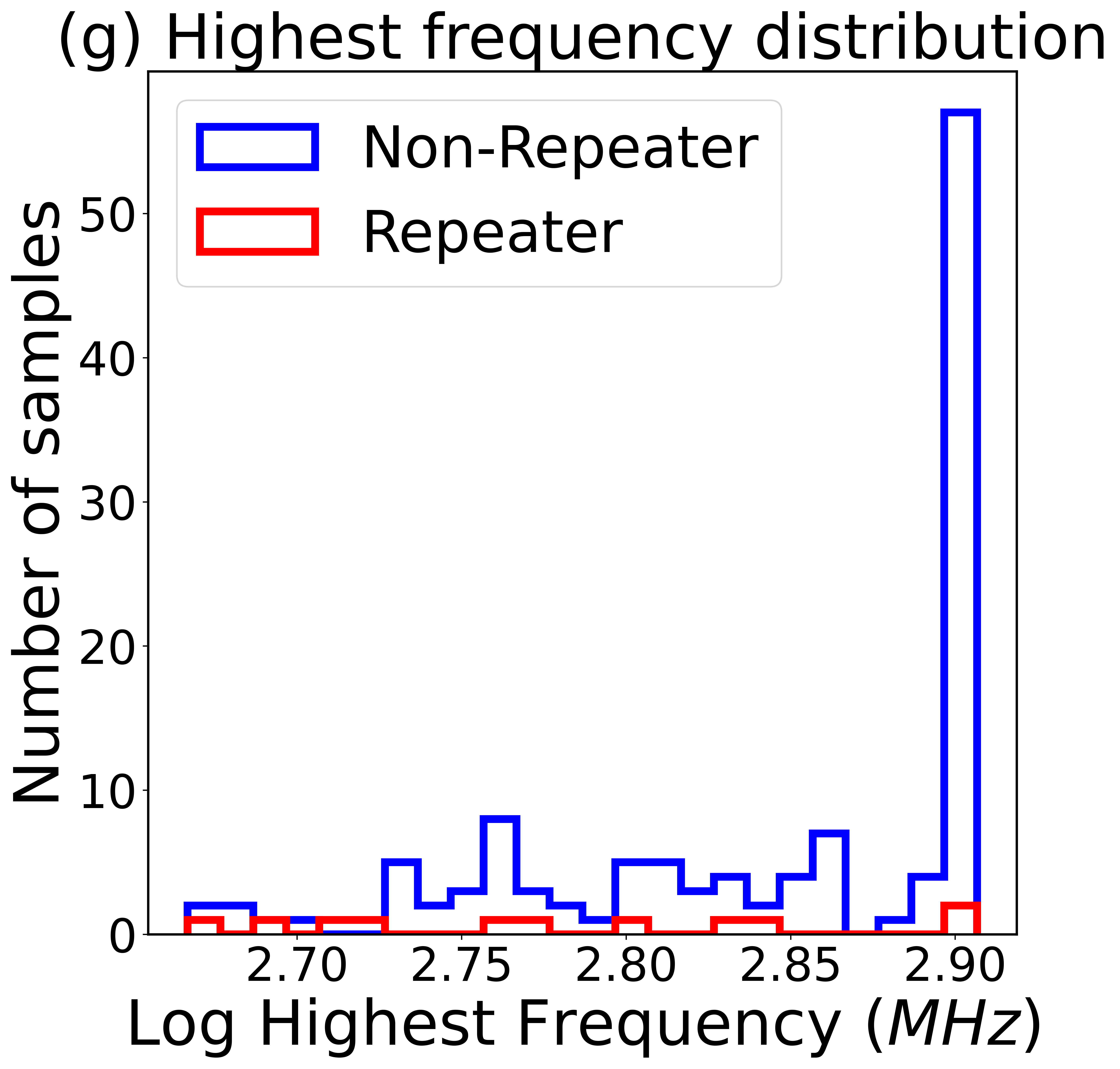}
    \includegraphics[width=0.3\textwidth]{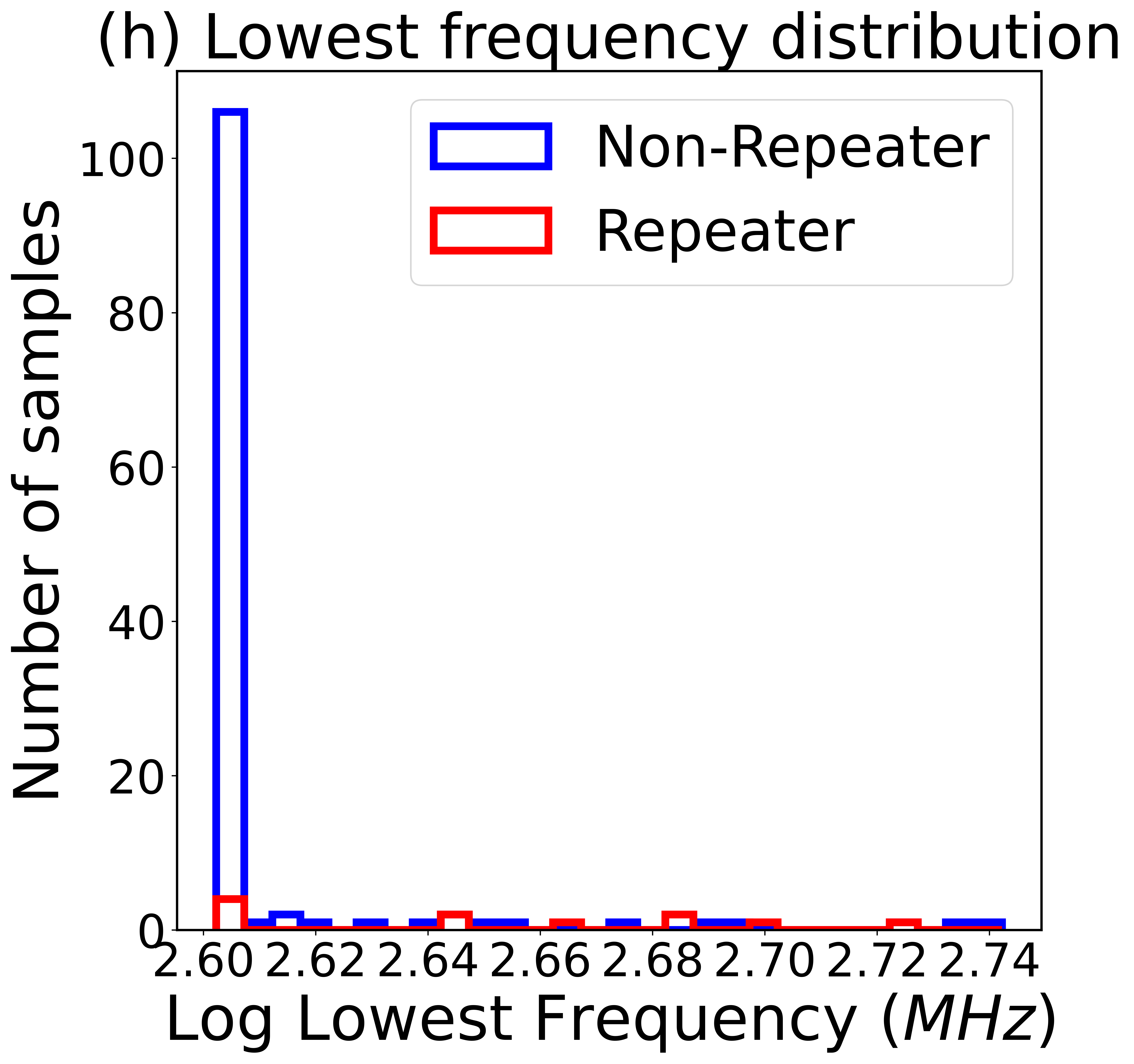}
    \includegraphics[width=0.3\textwidth]{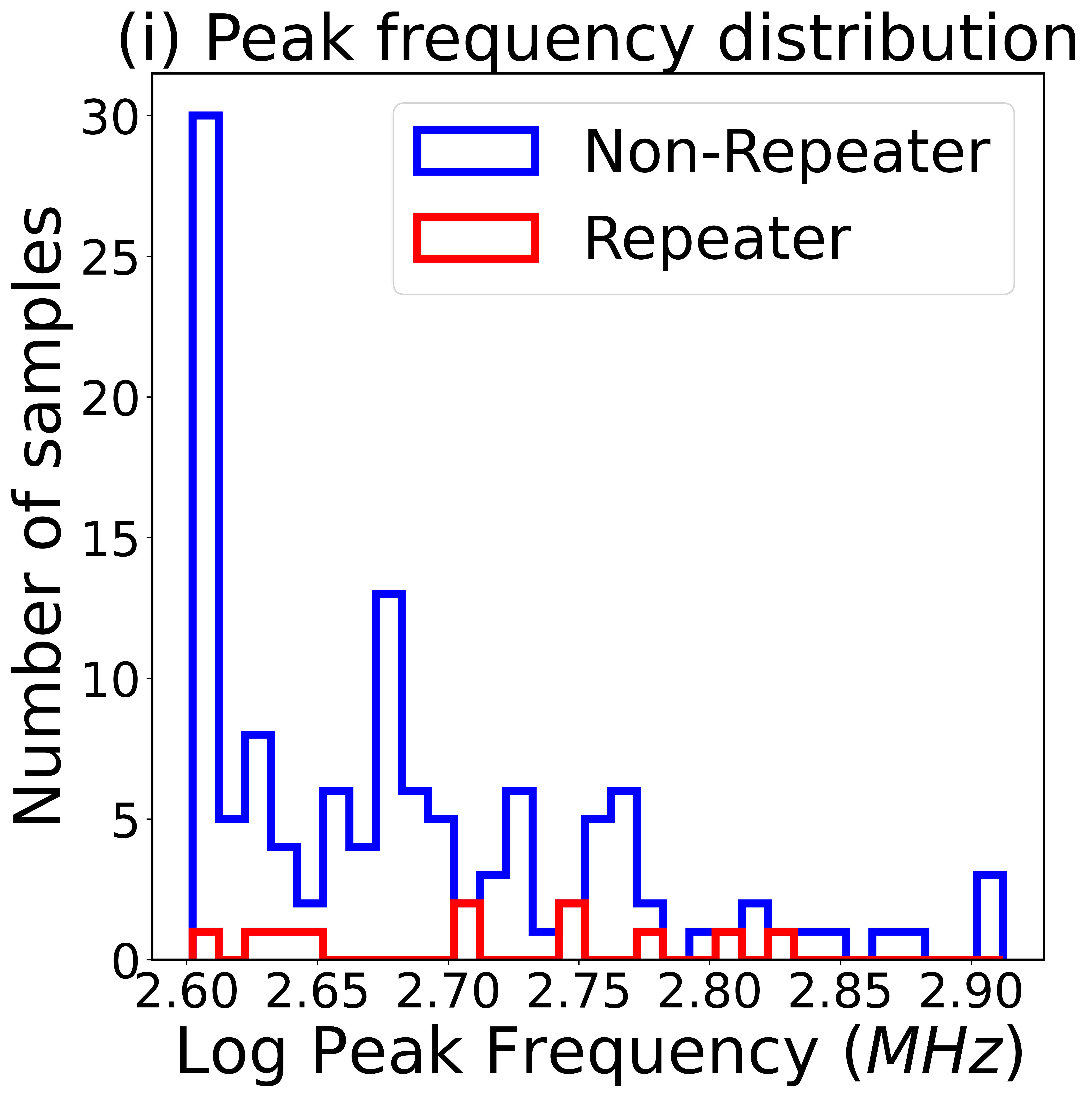}

    \includegraphics[width=0.3\textwidth]{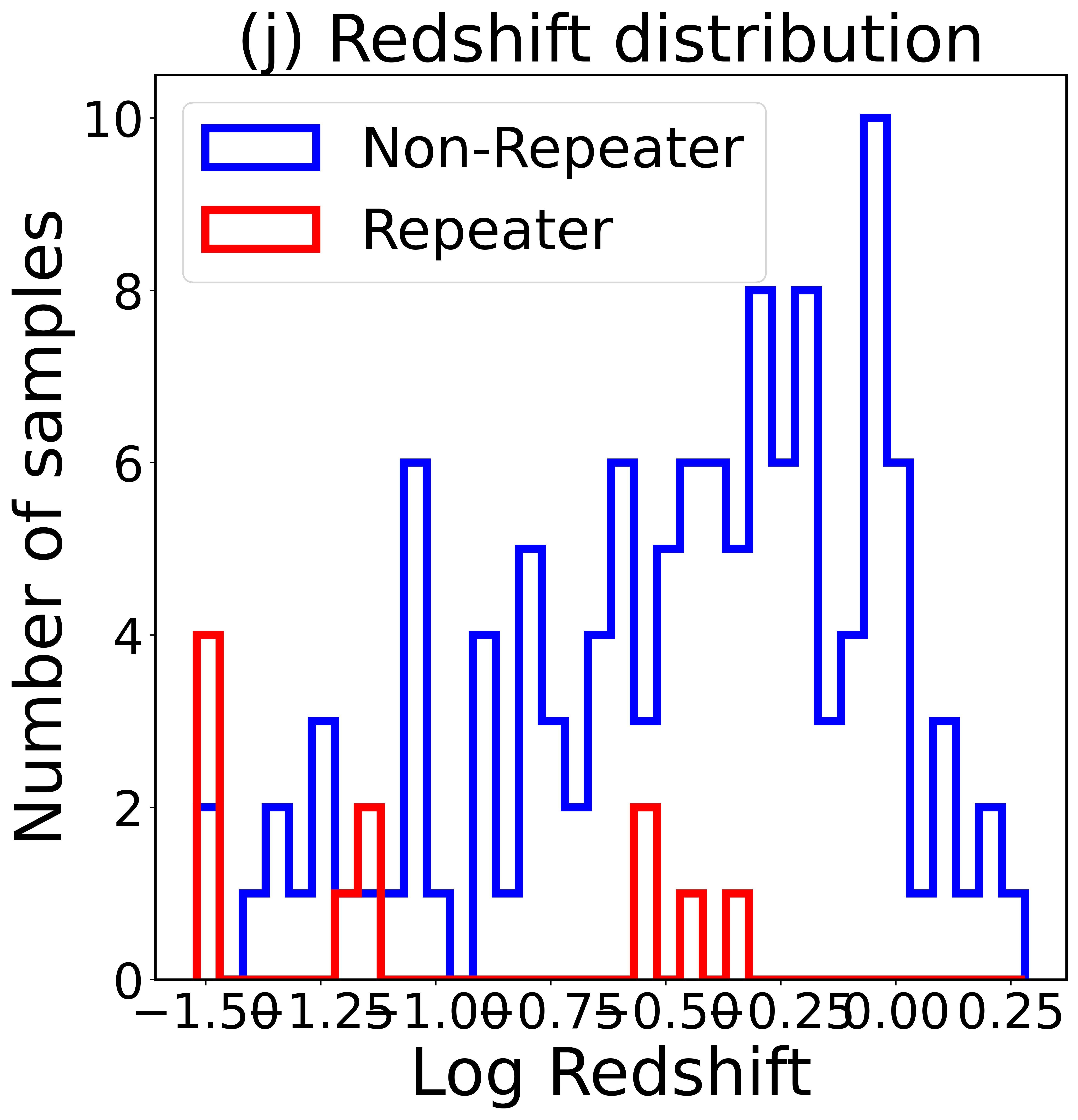}
    \includegraphics[width=0.3\textwidth]{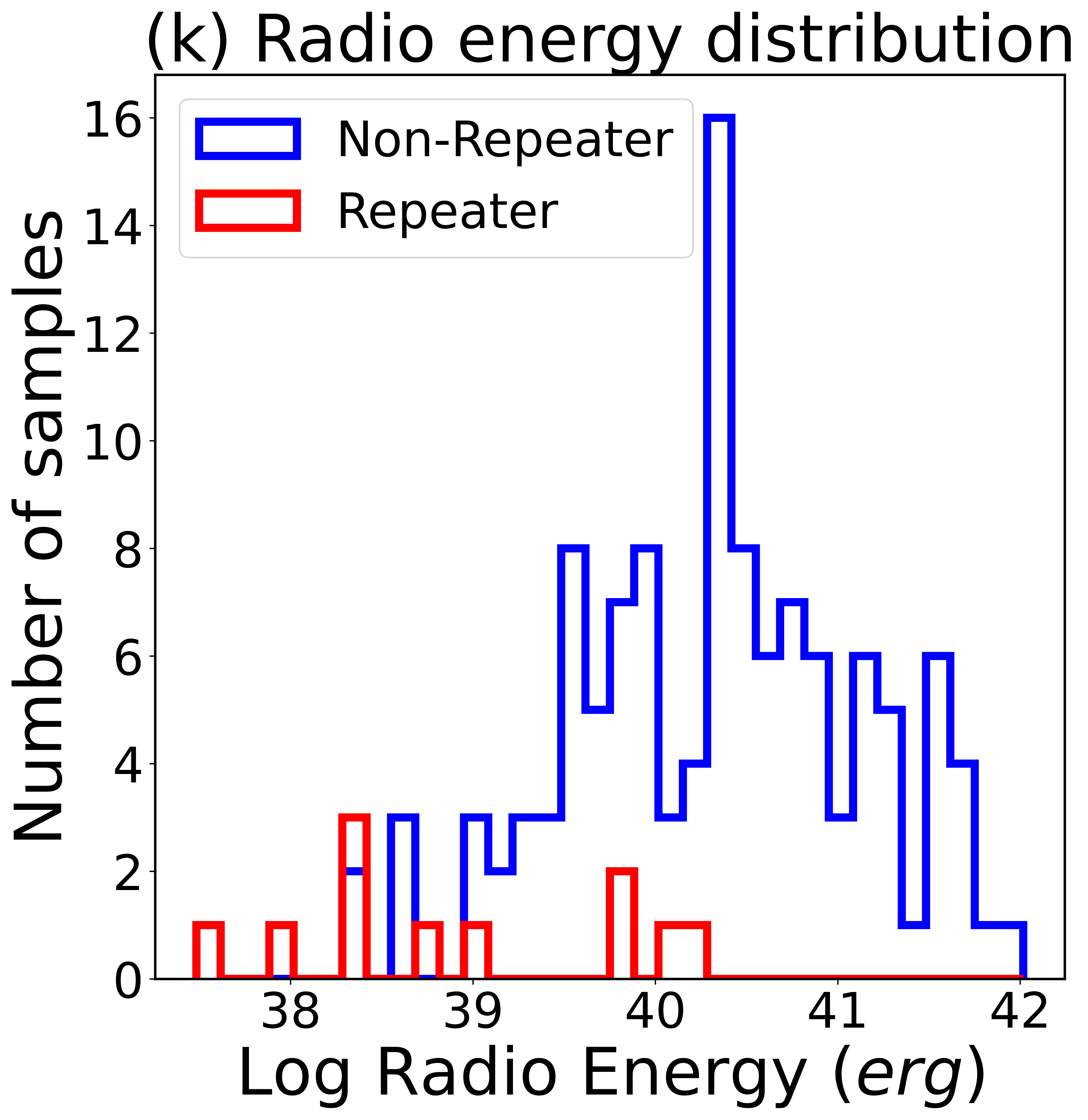}

    \caption{Distributions of both observed and model-dependent parameters for repeaters (red) and non-repeaters (blue), plotted after the sample selection described in Section \ref{sample}.}
    \label{fig:histogram}
\end{figure*}

\section{Machine learning model}\label{machine}
We employed an unsupervised machine learning approach to investigate the underlying structure of FRBs without labeled information. 
Specifically, we use Uniform Manifold Approximation and Projection (UMAP) for dimensionality reduction \citep{2018arXiv180203426M} and Hierarchical Density-Based Spatial Clustering of Applications with Noise (HDBSCAN) for clustering \citep{2019arXiv191102282M}. 

UMAP \citep{2018arXiv180203426M} is a nonlinear dimensionality reduction algorithm. 
It was developed based on topological data analysis and manifold theory. 
Further, UMAP has better visualization quality than t-distributed Stochastic Neighbor Embedding (t-SNE) \citep{maaten2008visualizing}, alongside advantages like faster runtime, better preservation of the global structure of the data, and the ability to handle larger datasets.
It is a general-purpose dimensionality reduction algorithm for machine learning because it does not have computational restrictions on embedding dimensions. 
It works on a solid theoretical foundation and mathematical framework, and is not derived with a task-focused objective function. 
This mathematical framework helps to minimize the cross-entropy between high and low-dimensional representations. 

We present the hyperparameters of UMAP that we use below. In the next section, we will explain how the hyperparameter values are selected.

$n\_neighbors$: It controls the balance between local and global structure in the data by determining the size of the local neighborhood used for manifold approximation; smaller values emphasize local structure and can lead to tighter grouping, while larger values preserve more global relationships \citep{2018arXiv180203426M}. 

$n\_components$: This hyperparameter represents the dimensionality of the embedding space. 
In this work, it was set to 2 for effective 2D visualization. 

$min\_dist$: This represents the closeness between the data points in high and low-dimensional space. 
It also controls the density of the low-dimensional embedding. 

Additionally, to ensure the reproducibility of the results, we fix the $random\_state$ hyperparameter as 1.
UMAP is a stochastic method that relies on randomness to approximate high-dimensional relationships and optimize the low-dimensional embedding. Therefore, setting a fixed $random\_state$ ensures that the results are reproducible across multiple runs.
Moreover, we use the cosine distance metric to measure the similarity between the data points. 
This metric is suitable for high-dimensional datasets \citep{2018arXiv180203426M}. 
Readers are referred to \cite{2018arXiv180203426M} for a detailed mathematical framework and description of UMAP's hyperparameters. 

HDBSCAN \citep{2019arXiv191102282M} is a clustering algorithm that identifies clusters based on density. 
This algorithm builds the cluster hierarchy tree and then uses stability measures to obtain the most significant groupings from the hierarchy. 
In density-based clustering, dense groups of points are separated by regions of lower density. 
The dense groups are identified as clusters. 
Groups falling below a specified density threshold level are classified as noise.

HDBSCAN is the advanced version of the Density-Based Spatial Clustering of Applications with Noise \citep[DBSCAN:][]{1996kddm.conf..226E, 2019arXiv191102282M}.
DBSCAN uses a pre-defined number of clusters in identifying clusters.
This leads to a significant variation in densities in clusters. 
Therefore, cluster identification does not guarantee sufficient data density in each cluster, wherein some cluster identifications could be less significant.
On the other hand, HDBSCAN does not rely on the pre-defined number of clusters.
It constructs a hierarchy of clusters across all possible densities above a certain density threshold.
For more details about HDBSCAN, we refer to \cite{2019arXiv191102282M}.

In this work, we employ the HDBSCAN hyperparameters listed below:

$min\_cluster\_size$: This determines the least number of samples needed for a cluster to emerge. 
It directly influences the granularity of the clustering. 
Smaller values allow detection of smaller, denser clusters, while larger values favor broader, more general groupings. 

$min\_sample$: It controls the sensitivity of the algorithm to noise and the definition of core points in a cluster. 

$cluster\_selection\_epsilon$: It sets a threshold for the minimum separation between clusters. The default value 0.1 was used in this work. 

$alpha$: This hyperparameter balances the influence of mutual reachability distance in the computation of the condensed tree. The default value 1.0 was used in this work. 

We systematically optimized the following hyperparameters in both UMAP and HDBSCAN: $n\_neighbors$, $min\_dist$, $min\_samples$, and $min\_cluster\_size$. The following section explains the optimization process in detail.

\subsection{Hyper parameter optimization}\label{hyper}
To optimize the hyperparameter, we use grid search by systematically evaluating the different combinations of hyperparameters. 
The considered hyperparameters in this search include $n\_neighbors$, $min\_cluster\_size$, $min\_dist$, and $min\_samples$. 
The $n\_neighbors$ ranges from 2 to 16. $min\_cluster\_size$ ranges from 3 to 10. 
$min\_dist$ ranges from 0.007 to 0.03, and $min\_samples$ ranges from 2 to 4.

The silhouette score \citep{rousseeuw1987silhouettes} and Davies-Bouldin score \citep{davies2009cluster} are the metrics employed to evaluate the clustering performance. 
While the silhouette score calculates the cohesion and separation of clusters, where the higher value demonstrates well-defined clusters, the Davies-Bouldin score measures the compactness and separation between the clusters, where lower values indicate better cluster performance.

The parameters $min\_dist$ in UMAP and $min\_sample$ in HDBSCAN are crucial for controlling how clusters are formed. Specifically, they determine the algorithm’s sensitivity to density, which is a key factor in identifying distinct groups. 
To systematically assess the clustering performance under various density combinations, we conducted a comprehensive grid search. This involved testing six different values (0.007, 0.008, 0.009, 0.01, 0.02, 0.03) for $min\_dist$ and three (2, 3, 4) for $min\_samples$. 
The six and three values include 18 combinations of the two parameters, where we also varied UMAP’s $n\_neighbors$ (ranging from 2 to 16) and HDBSCAN’s $min\_cluster\_size$ (ranging from 3 to 10).
For each of these parameter combinations, we calculated the silhouette score and the Davies-Bouldin score to quantitatively assess the clustering quality.

To visualize the results, we generated a series of plots for each of the 18 parameter combinations. 
In these plots, the silhouette score (or Davies-Bouldin score) was plotted as a function of $n\_neighbors$, with separate lines representing the variation for each $min\_cluster\_size$. 
This approach allowed us to identify the highest silhouette scores and lowest Davies-Bouldin scores, leading to the selection of the optimal hyperparameter combination for our dataset.
We adopt $min\_dist=0.01$ and $min\_samples=4$ because we found that these two values provide the highest (lowest) silhouette (Davies-Bouldin) scores in the grid search.

\FloatBarrier
\begin{figure}[ht]
    \centering
    \includegraphics[width=0.47\textwidth]{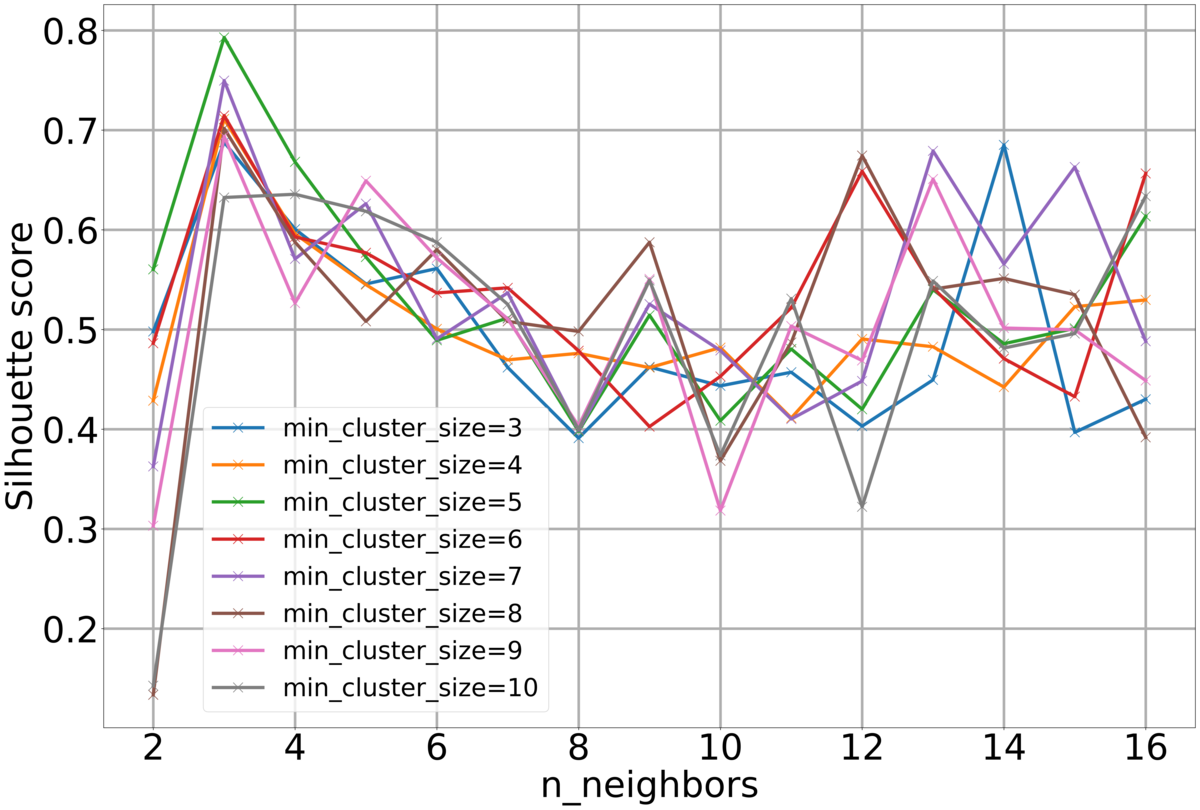}
    \caption{The silhouette score result of grid search. The $min\_cluster\_size=5$ reaches the maximum peak at $n\_neighbor=3$.
    The figure is shown with $min\_dist=0.01$ and $min\_samples=4$.
    }
    \label{fig:shilouette_score}
\end{figure}
\FloatBarrier

\FloatBarrier
\begin{figure}[ht]
    \centering
    \includegraphics[width=0.47\textwidth]{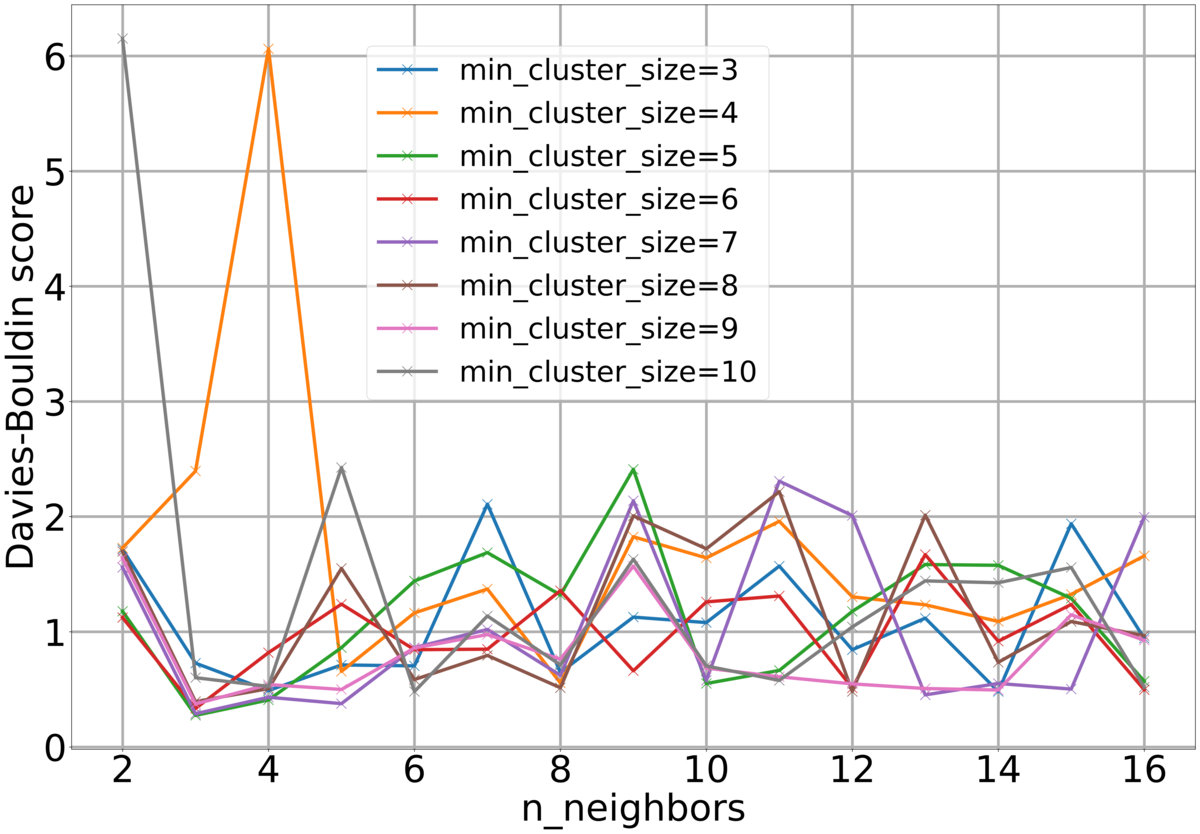}
    \caption{The daives-bouldin score result of grid search. The $min\_cluster\_size=5$ reaches the lower peak at $n\_neighbors=3$.
    The figure is shown with $min\_dist=0.01$ and $min\_samples=4$.
    }
    \label{fig:Davies_score}
\end{figure}
\FloatBarrier

The result of optimal hyperparameter is shown in Fig. \ref{fig:shilouette_score} and Fig. \ref{fig:Davies_score}. 
In Fig. \ref{fig:shilouette_score}, the grid search demonstrates that the $min\_cluster\_size$ of 5 achieves the highest Silhouette score of 0.792 at $n\_neighbors=4$, which means clusters are well separated and cohesive with a cohesion rate of 79.2\%. 
This score is obtained systematically for the combination of $min\_dist=0.01$ and $min\_samples=4$.
Similarly, in Fig. \ref{fig:Davies_score}, the best minimum Davies-Bouldin score of 0.273 is obtained for the same hyperparameter configuration, which indicates the optimal stability between well-cluster separation and compactness.

\subsection{Optimized Model configuration}\label{optimized}
The results of the hyperparameter optimization process are discussed in the previous section.
This configuration was selected by jointly considering the highest silhouette score and the lowest Davies-Bouldin score.  

The optimal hyperparameters of UMAP are
$n\_neighbors=3$, $min\_dist=0.01$, $n\_components=2$, $random\_state=1$, and chosen metric is cosine. 
The optimal hyperparameters of HDBSCAN are
$min\_cluster\_size=5$, $min\_sample=4$, $cluster\_selection\_epsilon=0.1$, and $alpha=1.0$.

According to \cite{2018arXiv180203426M}, using a smaller value for $n\_neighbors$ helps UMAP capture manifold structure accurately. 
In contrast, larger values can capture larger-scale manifold structures with a loss of fine details. 
In our testing, we tried values from $n\_neighbors=2$ to 16, and found that a smaller value of 3 provides the best results (Figs. \ref{fig:shilouette_score} and \ref{fig:Davies_score}). 
This indicates that our UMAP model finds denser structures. 
The $min\_dist$ directly influences the UMAP output. 
For this hyperparameter, a lower value indicates the potentially denser regions, and also collective manifold structures \citep{2018arXiv180203426M}. 

Overall, the optimized hyperparameter configurations enable UMAP to detect conjoint FRBs in the clusters, indicating the great similarity among the FRBs within each cluster. 
This is further supported by a high silhouette score of 0.792 and a lower value of the Davies-Bouldin score of 0.273. 

\subsection{Model Evaluation}\label{evaluation}
To assess the model performance of UMAP classification, we used k-fold cross-validation described in \cite{bishop2006pattern}. 
We use $k=6$.
Therefore, the repeater in the dataset is split into six different folds, where five folds are used for training while the remaining one fold is used for validation. 
This process is repeated until each fold serves as the validation fold once. Then we employed the \textit{F1} score \citep{2020arXiv201016061P} to calculate the accuracy. The \textit{F1} score is a metric that provides a balanced measure of classification performance of the model by combining the precision and recall. The \textit{F1} score metric is well-suited for datasets with imbalanced classes. For instance, our dataset has a larger number of non-repeaters than repeaters. Hence, we adopted the \textit{F1} score metric. A high \textit{F1} score indicates that a significant percentage of the positive class was accurately identified by the model. In this work, the positive class represents the repeaters. We used the \textit{F1} score to assess the performance of our model on the validation set of each fold, specifically concerning the ability to identify the repeaters. The average \textit{F1} score across all folds provides a robust estimate of the overall performance of the model. The \textit{F1} score calculation is provided below:

\FloatBarrier
\begin{figure}[tbp]
    \centering
    \includegraphics[width=0.47\textwidth]{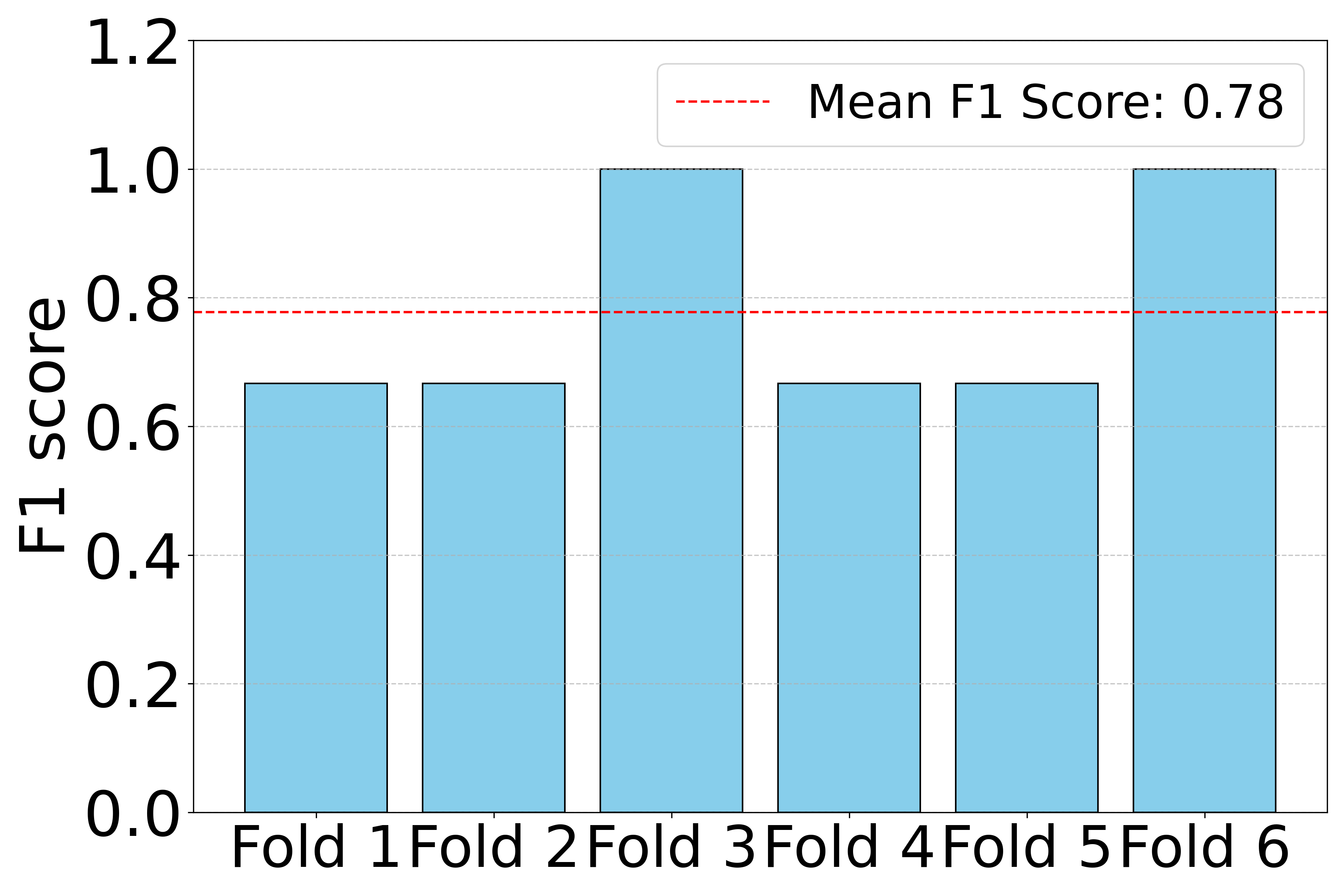}
    \caption{
    \textit{F1} Scores of the six-fold samples used for the cross-validation. 
    The mean \textit{F1} Score is presented by a red-dotted horizontal line.
    }
    \label{fig:cross_validation}
\end{figure}

\begin{equation}
\text{\textit{F1} score formula, \:\:}   \textit{F1} = 2 \times \frac{precision \times recall}{precision + recall},
\end{equation}

The expression for precision and recall is given in equations as follows. \\
\begin{equation}
    precision = \frac{TP}{TP + FP},
\label{precision}
\end{equation}

where TP (True positive) represents the repeaters that were correctly identified as repeaters. FP (False positive) represents the non-repeaters incorrectly identified as repeaters. 

\begin{equation}
    recall = \frac{TP}{TP + FN},
\end{equation}

where 
FN (False negative) denotes the repeaters that were incorrectly identified as non-repeaters. 


The \textit{F1} score results for each fold of the cross-validation, along with the mean score, are shown in Figure \ref{fig:cross_validation}. 
The mean score is 0.78, which demonstrates that the chosen UMAP configuration produces a meaningful representation of the FRB dataset and minimizes the risk of overfitting. 
This confirms that the UMAP delivers a robust low-dimensional representation of FRB data.

\section{Result}\label{result}
\subsection{UMAP training result with the Fold 1 sample}\label{umap-result}
Figure \ref{fig:machine_learning} shows the projection of unsupervised UMAP training for Fold 1 baseband FRB samples. 
The samples are grouped into three unique types, each illustrated by a different color. 
Specifically, non-repeating FRBs in training are shown in grey, repeating FRBs in training are in turquoise, and two repeating FRBs in the validation are in pink. 
Moreover, we include only repeating FRBs in the validation set because non-repeating FRBs cannot be validated due to the possible contamination from repeaters.

\FloatBarrier
\begin{figure}[tbp]
\centering
\includegraphics[width=1.0\linewidth]{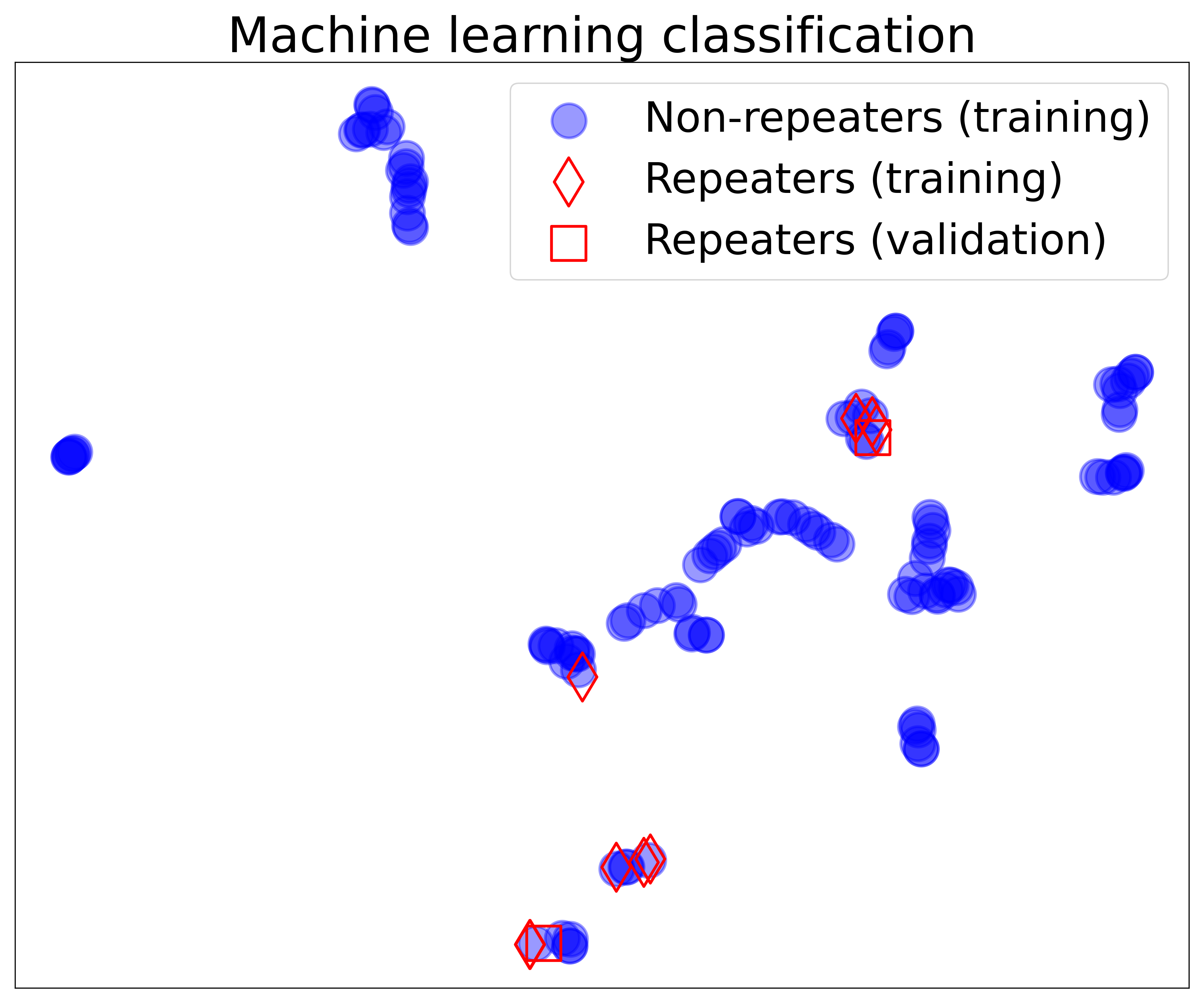}
\caption{
The unsupervised UMAP projection of the Fold 1 FRB samples.
The non-repeating (repeating) FRBs used for training are shown by grey (turquoise) dots.
The repeating FRBs used for the validation are shown by pink dots.
}
\label{fig:machine_learning}
\end{figure}

The UMAP training results show that the repeaters and non-repeaters form distinct clusters.
The validation repeaters are present inside the clusters where training repeaters dominate the cluster population. This indicates that the UMAP model captures a consistent structure in the dataset. Additionally, this consistency supports that the model is not overfitting.  
From UMAP training results, we notice that several non-repeating FRBs are closely present with known repeaters, particularly in clusters that are dominated by training repeaters (Fig. \ref{fig:machine_learning}). 
This result supports our initial hypothesis that some non-repeaters may be repeaters. 
These non-repeaters have not been detected more than once during FRB observations with the CHIME/FRB instrument. 
More importantly, our methodology has successfully recognized these mixed non-repeaters as potential FRB repeater candidates, strengthening the reliability of our approach.
On the other hand, only one training repeater appeared outside these repeater-dominated clusters nearly negligible number compared to the total training repeaters. 

\subsection{Identifying 13 clusters and the FRB repeater candidates with the entire sample} \label{identifying}
UMAP was trained on the entire dataset, and its output was subsequently used for cluster identification with HDBSCAN. 
Fig. \ref{fig:HDBSCAN_image} shows 12 clusters and 1 noise cluster identified by HDBSCAN.
Each cluster is shown by each color.

\begin{figure}[tbp]
    \centering
    \includegraphics[width=0.48\textwidth]{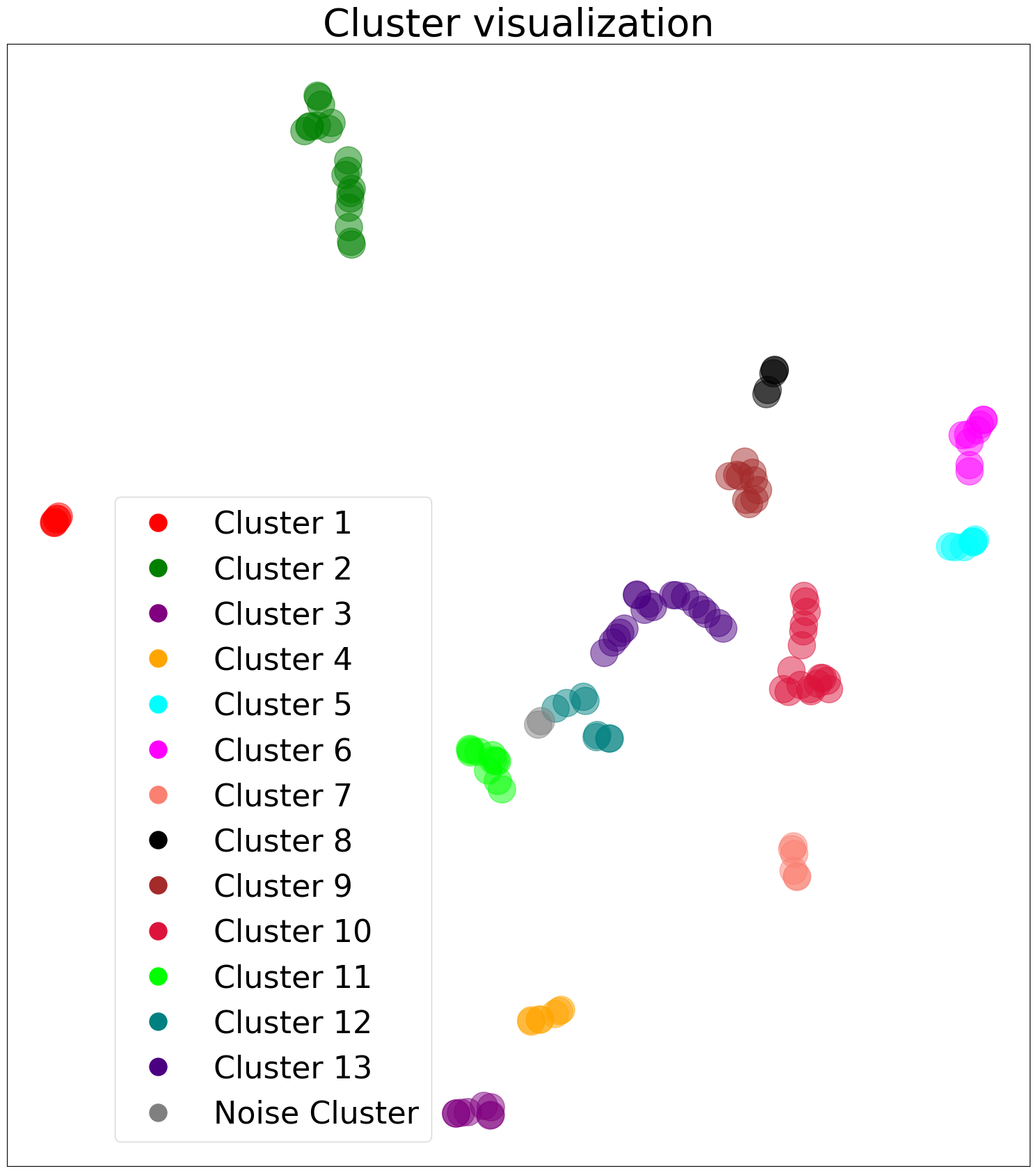}
    \caption{
    UMAP projection of the dataset colored by HDBSCAN cluster assignments. A total of 13 dense clusters and one noise cluster were identified by HDBSCAN, each represented by a distinct color and labeled as cluster 1 through cluster 13. The noise cluster is shown in grey. 
    }
    \label{fig:HDBSCAN_image}
\end{figure}

To assess the implementation of UMAP and identify the potential repeater candidates, we applied a repeater threshold.
The repeater threshold is the threshold applied for the fraction of repeaters in each cluster to identify repeater clusters.
\citet{2022MNRAS.509.1227C} adopted a very low repeater threshold of 10\%, above which a cluster is identified as a repeater cluster, involving the CHIME intensity catalog.
In contrast, we aim to use the maximum threshold as possible. 
The higher threshold indicates the larger number of repeaters within the repeater cluster. In this way, we can identify more reliable and suitable repeater candidates that have more similar physical properties to repeaters. Hence, we tried to maximize the threshold.
Therefore, we propose a process that employs the precision (Equation \ref{precision}) as a function of the repeater threshold, namely completeness-guided threshold selection.
TP values are calculated based on repeater thresholds ranging from 30\% to 40\%.


Fig. \ref{fig:threshold} illustrates the result of this process, where the model performance remains at approximately 90\% up to a threshold of 37\%, after which it declines gradually. 
So, based on these findings, we have established the repeater threshold at 37\%. 
We have increased the threshold level by more than three times from the previous study \citep{2022MNRAS.509.1227C}. 
This adjustment ensures the identification of more suitable repeater candidates.

\begin{figure}[tbp]
    \centering
    \includegraphics[width=0.5\textwidth]{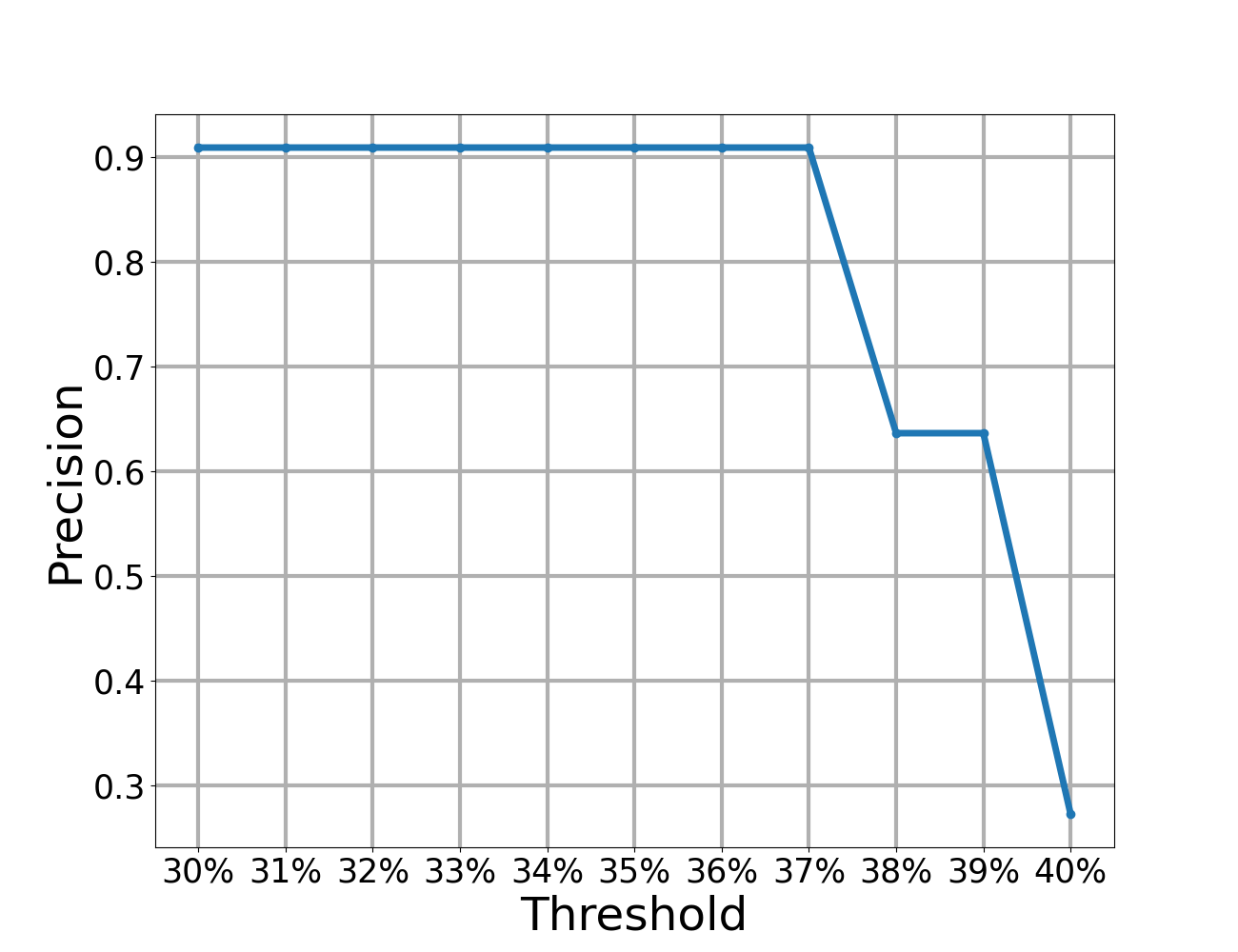}
    \caption{
    Precision as a function of the repeater threshold.
    The model performance is steady up to the threshold level of 37\%.}
    \label{fig:threshold}
\end{figure}

We do not assess the false-negative (non-repeaters in repeater clusters) accuracy because these metrics necessitate the availability of ground truth for non-repeaters, which is not yet confirmed in FRB studies.
Fig. \ref{fig:cluster_image} shows repeater and non-repeater clusters highlighted in different colors and markers, based on the 37\% repeater threshold.
Three clusters are identified as repeater clusters, 
while the remaining 10 clusters are categorized as non-repeater clusters, and 1 noise cluster was identified. 
They are labeled as Repeater cluster 1 -3,  Non-repeater cluster 1- 10, and noise cluster in grey color (see Fig. \ref{fig:cluster_image}).
Brief insights of each group are summarized in Table \ref{table:1}. 


On the other hand, one training repeater (FRB 20190621A) lies in Non-repeater Cluster 8, which is away from the repeater cluster (Fig. \ref{fig:machine_learning}).
We hypothesize that this outlier repeater FRB may be due to the higher repeater threshold adopted in this work (37\%). 
With this high threshold, a statistically less significant repeater cluster is not identified as a repeater cluster.
Therefore, Non-repeater Cluster 8 could be classified as a repeater cluster only when a low repeater threshold is adopted.
However, we do not consider this cluster as a repeater cluster in the following analysis.

\begin{table}[tbp]
\centering
\caption{Number of samples in each cluster}
\label{table:1}
\resizebox{\columnwidth}{!}{%
\begin{tabular}{lccc}
\toprule
\textbf{Cluster Name} & \textbf{Total} & \textbf{Confirmed Repeater} & \textbf{Candidate} \\
\midrule
Repeater Cluster 1      & 8  & 3 & 5 \\
Repeater Cluster 2      & 7  & 3 & 4 \\
Repeater Cluster 3      & 10 & 4 & 6 \\
Non-Repeater Cluster 1  & 6  & 0 & 0 \\
Non-Repeater Cluster 2  & 19 & 0 & 0 \\
Non-Repeater Cluster 3  & 7  & 0 & 0 \\
Non-Repeater Cluster 4  & 9  & 0 & 0 \\
Non-Repeater Cluster 5  & 6  & 0 & 0 \\
Non-Repeater Cluster 6  & 5  & 0 & 0 \\
Non-Repeater Cluster 7  & 17 & 0 & 0 \\
Non-Repeater Cluster 8  & 11 & 1 & 0 \\
Non-Repeater Cluster 9  & 8 & 0 & 0 \\
Non-Repeater Cluster 10 & 18 & 0 & 0 \\
Noise cluster & 2 & 0 & 0 \\
\bottomrule
\end{tabular}
}
\end{table}

\FloatBarrier
\begin{figure}[tbp]
\centering
\includegraphics[width=0.9\linewidth]{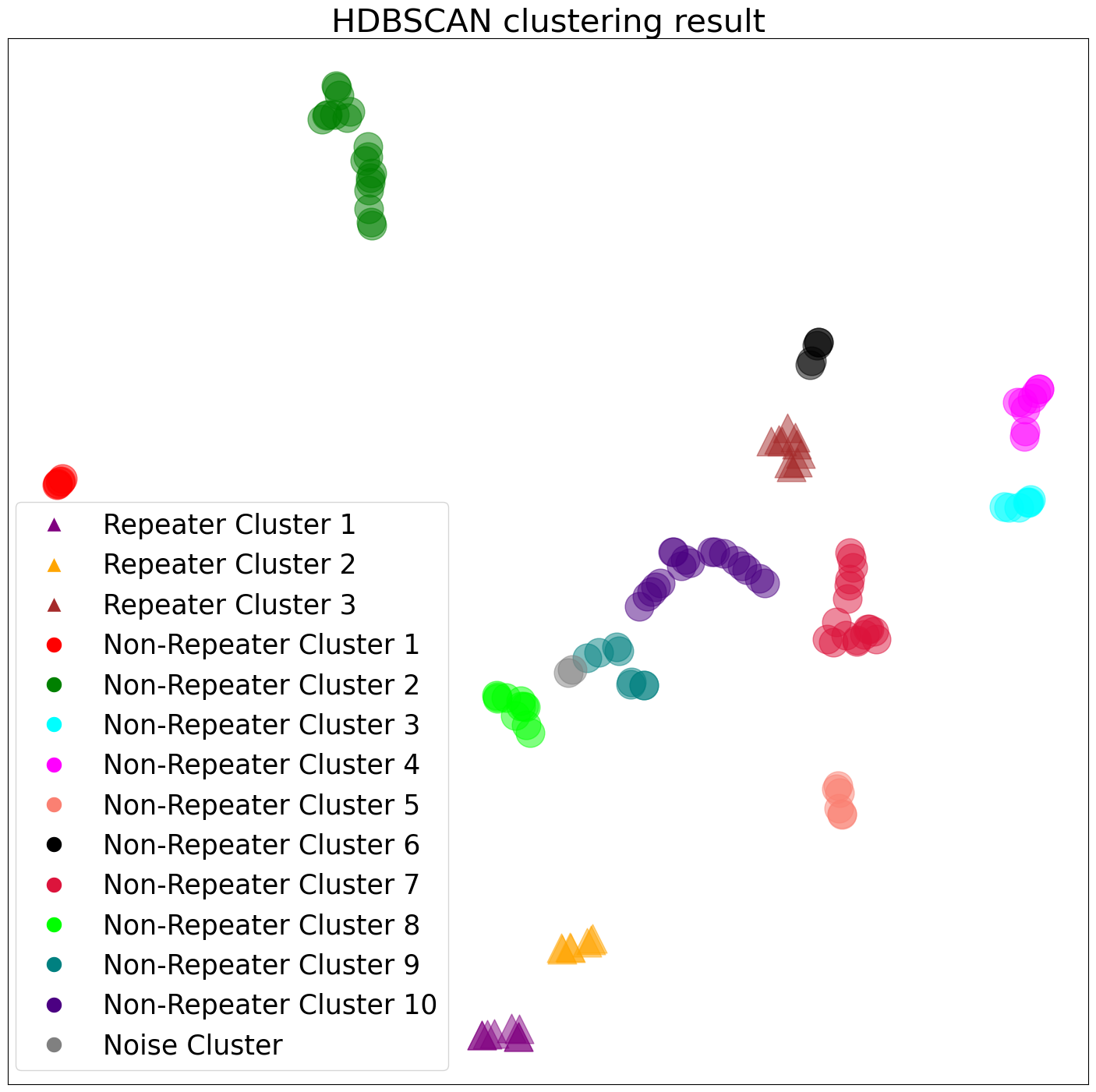}
\caption{The HDBSCAN algorithm yields a well-defined clustering of the projected FRB samples, resulting in 13 distinct clusters. Among these, three clusters are identified as associated with repeating FRBs and are designated as repeater clusters 1-3. The remaining clusters, corresponding to non-repeating FRBs, are labeled as non-repeater clusters 1-10. The noise cluster is shown in grey.}
\label{fig:cluster_image}
\end{figure}

The non-repeating FRBs in the repeater clusters are considered repeater candidates.
We plot the identified FRB repeater candidates along with repeaters and non-repeaters in Fig. \ref{fig:candidate_image}. 
Our technique efficiently gathers non-repeaters whose latent features are similar to those of the repeaters. 
As shown in Fig. \ref{fig:candidate_image}, we identify 15 repeater source candidates from a total of 122 non-repeater sources, representing the possible repeater fraction of 12.3\% in the non-repeater sample. 
The identified repeater candidates are listed in Table \ref{table:2}.

\begin{figure}[tbp]
\centering
\includegraphics[width=0.9\linewidth]{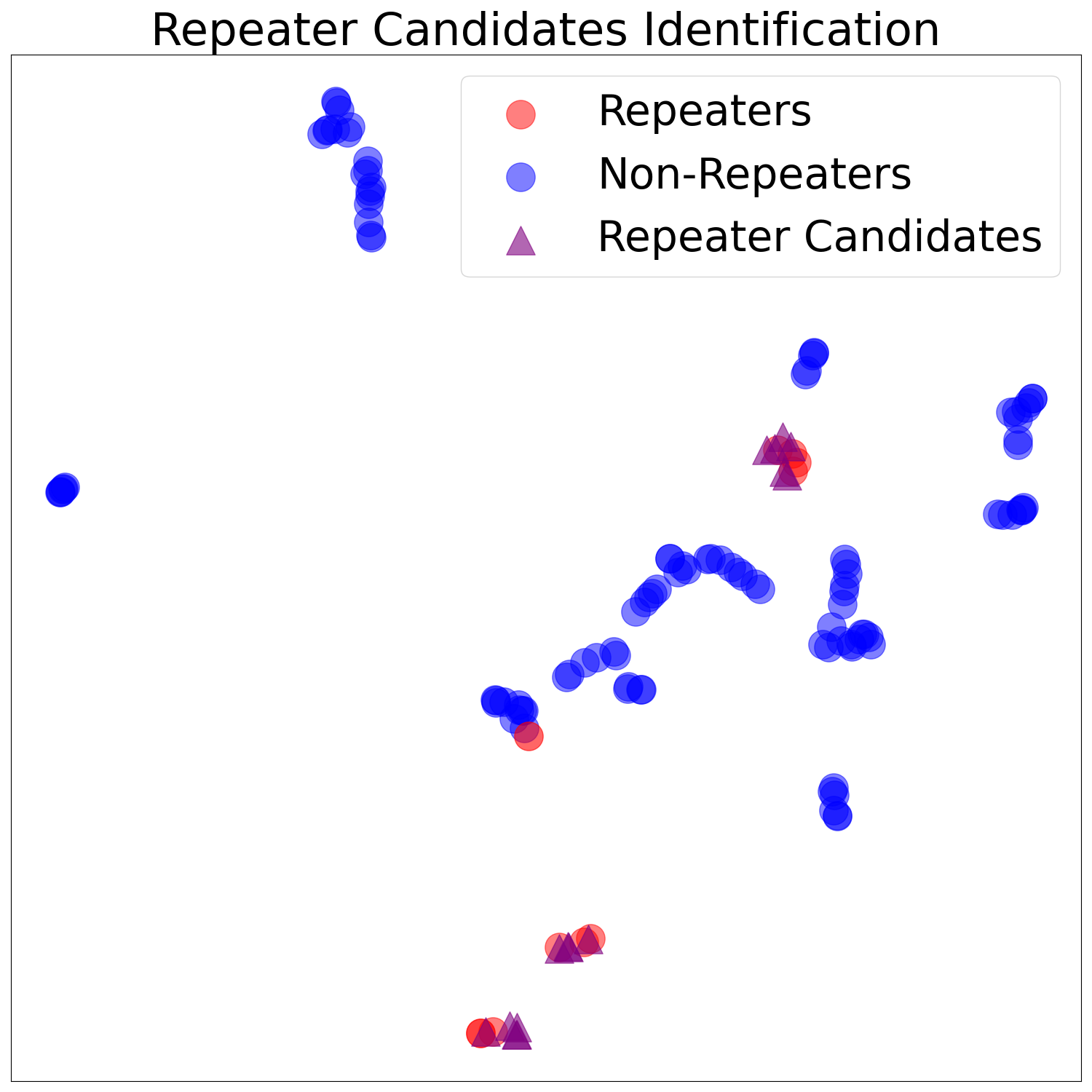}
\caption{The non-repeating FRBs embedded within the repeater clusters are classified as FRB repeater candidates and are indicated in blue.}
\label{fig:candidate_image}
\end{figure}

\begin{deluxetable*}{lccccccccccc}
\tablecaption{The list of identified FRB repeater candidates\label{table:2}}
\tabletypesize{\scriptsize}
\tablewidth{0pt}
\tablehead{
\colhead{FRB Name} & \colhead{bc width} & \colhead{sp idx} & \colhead{High freq} & \colhead{Flux} & \colhead{sp run} & \colhead{Radio Energy} & \colhead{$z$} & \colhead{Low freq} & \colhead{scat time} & \colhead{Fluence} & \colhead{Peak freq} \\
\colhead{} & \colhead{(s)} & \colhead{} & \colhead{(MHz)} & \colhead{(Jy)} & \colhead{} & \colhead{log$_{10}$(erg)} & \colhead{} & \colhead{(MHz)} & \colhead{(s)} & \colhead{(Jy ms)} & \colhead{(MHz)}
}
\startdata
FRB20181221A & 0.0079 & 62.1 & 583.3 & 3.3 & -128.0 & 39.7944 & 0.2079 & 446.1 & 0.00138 & 15.0 & 510.1 \\
FRB20181222E & 0.041 & 5.13 & 639.5 & 8.7 & -19.9 & 39.9747 & 0.1818 & 400.2 & 0.00084 & 30.0 & 455.2 \\
FRB20181228B & 0.0092 & 59.3 & 471.8 & 0.4 & -353.0 & 39.5656 & 0.4649 & 401.5 & 0.0011 & 1.67 & 435.2 \\
FRB20181231B & 0.0021 & 59.6 & 800.0 & 24.0 & -60.0 & 39.1884 & 0.0552 & 540.6 & 0.00134 & 56.0 & 657.7 \\
FRB20190102A & 0.011 & 28.9 & 595.5 & 15.0 & -67.8 & 41.5679 & 0.5977 & 411.9 & 0.00103 & 10.0 & 495.2 \\
FRB20190110C & 0.0046 & 24.5 & 477.7 & 3.4 & -186.0 & 38.6055 & 0.0937 & 400.2 & 0.00063 & 5.3 & 427.4 \\
FRB20190130B & 0.0038 & 55.4 & 553.6 & 13.0 & -140.8 & 41.3088 & 0.9004 & 428.6 & 0.00056 & 24.0 & 487.1 \\
FRB20190203A & 0.0074 & 25.0 & 563.4 & 12.0 & -75.0 & 40.5813 & 0.3033 & 400.2 & 0.00082 & 42.0 & 472.9 \\
FRB20190213D & 0.01 & 26.2 & 800.2 & 3.7 & -25.3 & 41.3359 & 1.0458 & 496.6 & 0.00233 & 19.0 & 671.4 \\
FRB20190430C & 0.0034 & 48.7 & 800.2 & 4.0 & -48.8 & 39.6831 & 0.2352 & 530.6 & 0.00083 & 8.6 & 659.3 \\
FRB20190519E & 0.00037 & 2.0 & 800.2 & 3.3 & 4.1 & 39.5876 & 0.6110 & 551.0 & 4e-05 & 1.5 & 800.2 \\
FRB20190609A & 0.01 & 62.4 & 683.4 & 16.0 & -84.0 & 40.0034 & 0.1695 & 491.0 & 0.0004 & 37.0 & 579.3 \\
FRB20190609C & 0.0023 & 15.2 & 481.3 & 3.0 & -138.0 & 39.3700 & 0.2456 & 400.2 & 3e-05 & 4.1 & 422.9 \\
FRB20190629A & 0.0059 & 24.7 & 733.6 & 6.8 & -35.3 & 40.6017 & 0.4062 & 440.1 & 0.0014 & 24.0 & 568.2 \\
FRB20190701C & 0.0039 & 46.2 & 495.5 & 15.0 & -211.0 & 41.1939 & 0.8433 & 402.2 & 0.00041 & 21.0 & 446.4 \\
\enddata
\tablecomments{
\textbf{bc width} = burst duration (s), \textbf{sp idx} = spectral index, \textbf{High/Low/Peak freq} = frequencies (MHz), \textbf{sp run} = spectral running, \textbf{z} = redshift, \textbf{scat time} = scattering time (s), \textbf{Flux} = in Jy, \textbf{Fluence} = in Jy$\cdot$ms.
}
\end{deluxetable*}

Considering the 11 original repeaters into account, the UMAP model anticipates an FRB repeater fraction of (11+15)/(11+122) = 19.5\%. 
Previously, only approximately 5\% of FRBs had been observed to be repeated \citep{2021ApJS..257...59C}, with a small extended estimate of 8\% reported in \cite{2024ApJ...969..145C}. 
Our findings propose a substantially larger repeater population, necessitating follow-up observations for confirmation. 

\section{Discussion}\label{discussion}
\subsection{Feature importance}\label{feature}
In our research, we exploit 9 observational parameters and 2 model-based parameters to train the UMAP model in an unsupervised learning. 
As said in Section \ref{result}, our approach provides a classification, successfully revealing FRB repeater candidates. 
To further figure out the contribution of each parameter to the model's performance, we conducted a feature importance analysis. 
Specifically, we accessed the permutation feature importance method, an extensively used model interpretation technique \citep{altmann2010permutation}. 
This approach involves two key steps. 
For a given feature, the values of the feature are randomly swapped across the repeater samples, keeping the values of the other features unchanged. For each feature used for shuffling, the model performance is calculated after this shuffling process. If the model performance is increased after this shuffling process, it means that the feature used for shuffling is not important. If the model performance decreases after this shuffling process, it means the feature is important. The shuffling process effectively breaks the association between the feature and the model’s prediction.
Second, the change in model performance is measured after the shuffling process. A substantial decrease in performance indicates that the feature is important for the model, whereas little or no decrease suggests that the feature is less important for the model. 


\FloatBarrier
\begin{figure}[tbp]
\centering
\includegraphics[width=0.47\textwidth]{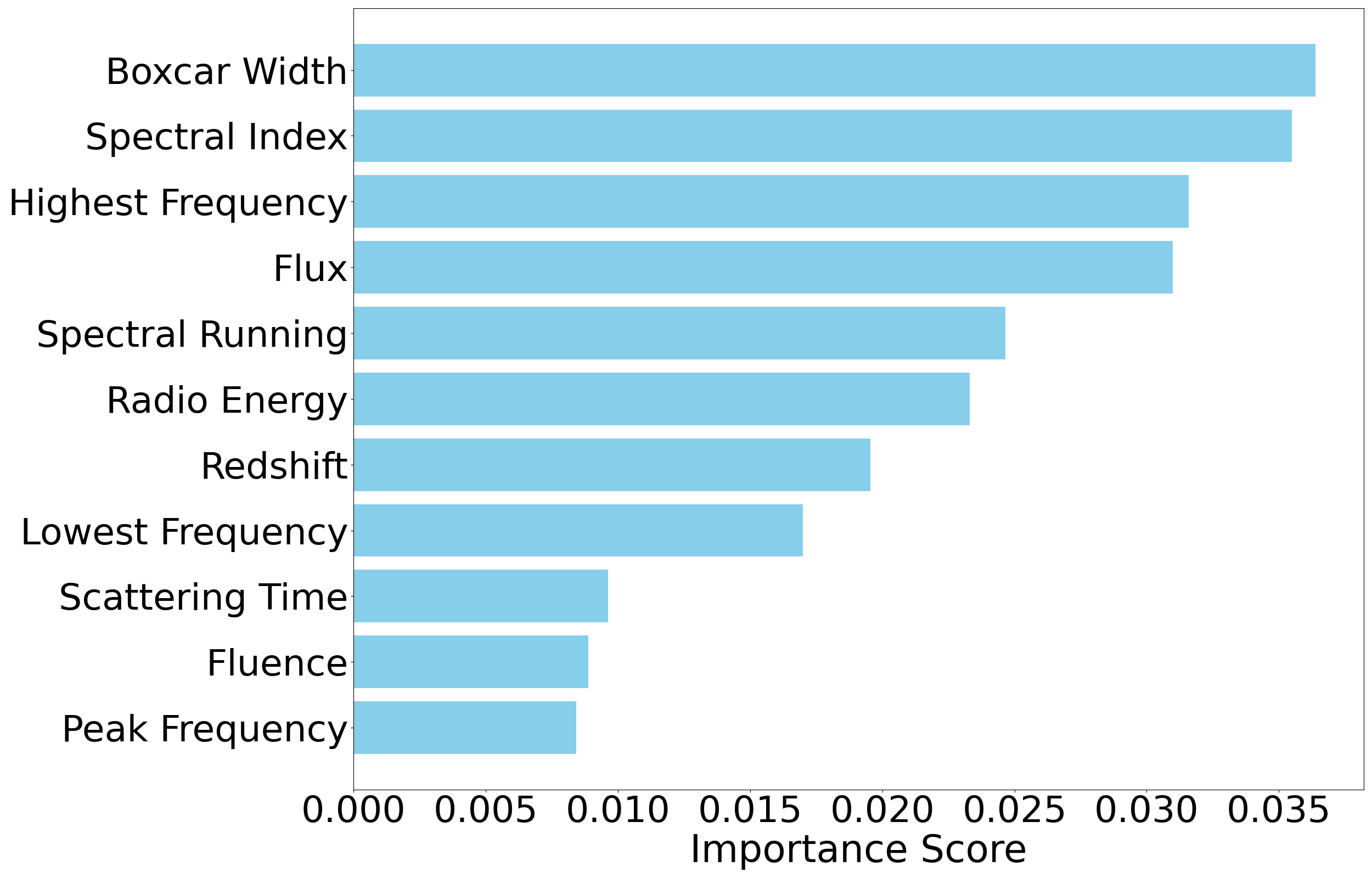}
\caption{The result of permutation feature importance for the optimized UMAP model.}
\label{fig:feature_importance}
\end{figure}

The outcome of the permutation feature importance analysis is shown in Fig. \ref{fig:feature_importance}, where the performance metric is the precision of repeaters, as defined in Equation \eqref{precision} in Section \ref{identifying}. 
Our findings indicate that pulse duration (Boxcar width) is the most important feature for FRB classification, with peak frequency contributing the least to the model's performance. 

In this work, we focused on an important feature for further analysis. Moreover, it is evident that multiple features collectively contribute to the machine learning classification outcome (see, Fig. \ref{fig:feature_importance}). While 'Boxcar Width' and 'Spectral Index' show the highest importance scores, their values (approximately 0.035) are not drastically higher than those of other significant features such as 'Highest Frequency' (around 0.032) and 'Flux' (around 0.031). This relatively even distribution among the top-ranked features indicates that no single feature overwhelmingly dominates the machine learning results.  


\subsection{Comparison with \cite{2022MNRAS.509.1227C}} \label{compare}
\cite{2022MNRAS.509.1227C} have identified 188 repeater candidates from the CHIME intensity catalog. 
In this work, we identify 15 repeater candidates from the CHIME baseband catalog. 
The CHIME baseband catalog is an updated version of 140 FRB samples from the CHIME intensity catalog. 
In other words, the CHIME intensity catalog also contains the baseband samples. 
In this scenario, some FRBs can be classified as repeater candidates by both this work and \cite{2022MNRAS.509.1227C}. 
The common repeater candidates in both work indicate that the possibility of repeater nature for these FRBs is high.  

\FloatBarrier
\begin{figure}[tbp]
\centering
\includegraphics[width=0.9\linewidth]{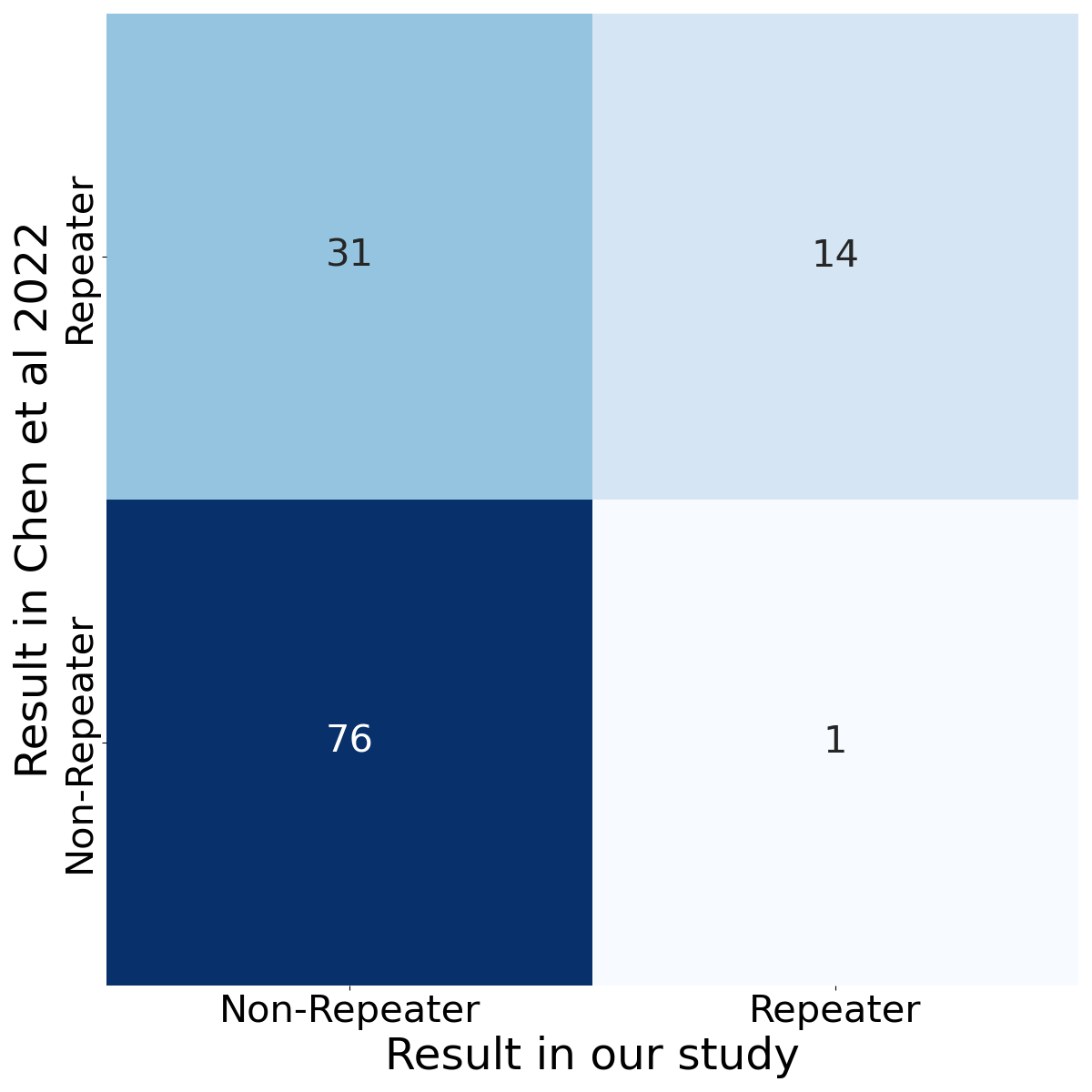}
\caption{The heatmap shows the agreement rate between this work and \cite{2022MNRAS.509.1227C}, highlighting the following groups: 14 common repeater candidates, 76 common non-repeaters, 1 newly identified repeater candidate in our research.}
\label{fig:heat_map}
\end{figure}

Additionally, we present the distribution of agreement and disagreement of classification results between this work and \cite{2022MNRAS.509.1227C} using a confusion matrix in Fig. \ref{fig:heat_map}. 
This confusion matrix exhibits the relationship between these two classification results, displaying areas of powerful concurrence as well as instances of classification divergence.
In Fig. \ref{fig:heat_map}, 14 FRBs are commonly predicted as repeater candidates both in \cite{2022MNRAS.509.1227C} and this work.
We found one new repeater candidate. 
31 FRBs are classified as repeater candidates in \cite{2022MNRAS.509.1227C}, but they are not repeater candidates in this work (hereafter, these samples are mentioned as conflict samples).
76 FRBs commonly remain non-repeaters in both \cite{2022MNRAS.509.1227C} and this work.
The 14 common repeater candidates, one new repeater candidate, 31 conflict samples, and the 76 non-repeaters are discussed in Sections \ref{14repeaters}, \ref{duration}, \ref{conflict}, and \ref{non-repeaters}, respectively. 

\subsubsection{14 common repeater candidates}
\label{14repeaters}
The strong agreement on the 14 common repeater candidates shows that these \citep[this work and][]{2022MNRAS.509.1227C} FRB classifications are reliable. 
Even though the two models use different feature hierarchies to make their decisions, these 14 FRBs are still identified as repeater candidates.
Additionally, compared to \cite{2022MNRAS.509.1227C}, our dataset benefits from enhanced measurements of duration, flux, and fluence, yet these candidates persistently stick out across models. 
A recent study conducted an empirical analysis of 36 non-repeating FRBs, as reported in \cite{2025MNRAS.540.3709U}.
Their samples included FRB 20181221A, FRB 20181228B, and FRB 20190102A for follow-up observation using the Five-Hundred-meter Spherical Radio Telescope \citep[FAST;][]{2011IJMPD..20..989N}. 
These FRBs were chosen as potential repeater candidates from the repeater candidate list of \cite{2022MNRAS.509.1227C}. Notably, all three FRBs are also identified as repeater candidates in our work. 
However, there is no FRB detection in the follow-up observations by \cite{2025MNRAS.540.3709U}. This might be due to their very short exposure time (10 min) on each source. The FRB 20190110C was also recently confirmed as a repeating source by the CHIME/FRB collaboration \citep{2025ApJ...982..154N}. Interestingly, this particular FRB was also identified as a common repeater candidate in both our study and \cite{2022MNRAS.509.1227C}. 

In summary, 15 repeater candidates were identified in this work. Among them, 14 were also listed as repeater candidates in \cite{2022MNRAS.509.1227C}. One of these has been confirmed as a repeater. So, 13 common candidates and one new candidate from our study remain unconfirmed. Based on the evidence, we strongly recommend conducting follow-up observations on these 14 candidates to confirm their repeating nature.

\subsubsection{A new repeater candidate}\label{duration}
In our research, we found one new repeater candidate and 14 common candidates with \cite{2022MNRAS.509.1227C}, as explained in section \ref{compare}. In distinction to non-repeaters, repeating FRBs typically have wider durations, as evidenced by their broader band-averaged temporal profiles but narrower frequency ranges \citep{2021ApJ...923....1P}. According to the feature importance of our study, duration is the most significant feature, so we compared the baseband and intensity pulse profiles of our repeater candidates. 

\FloatBarrier
\begin{figure}[tbp]
\centering
\includegraphics[width=0.9\linewidth]{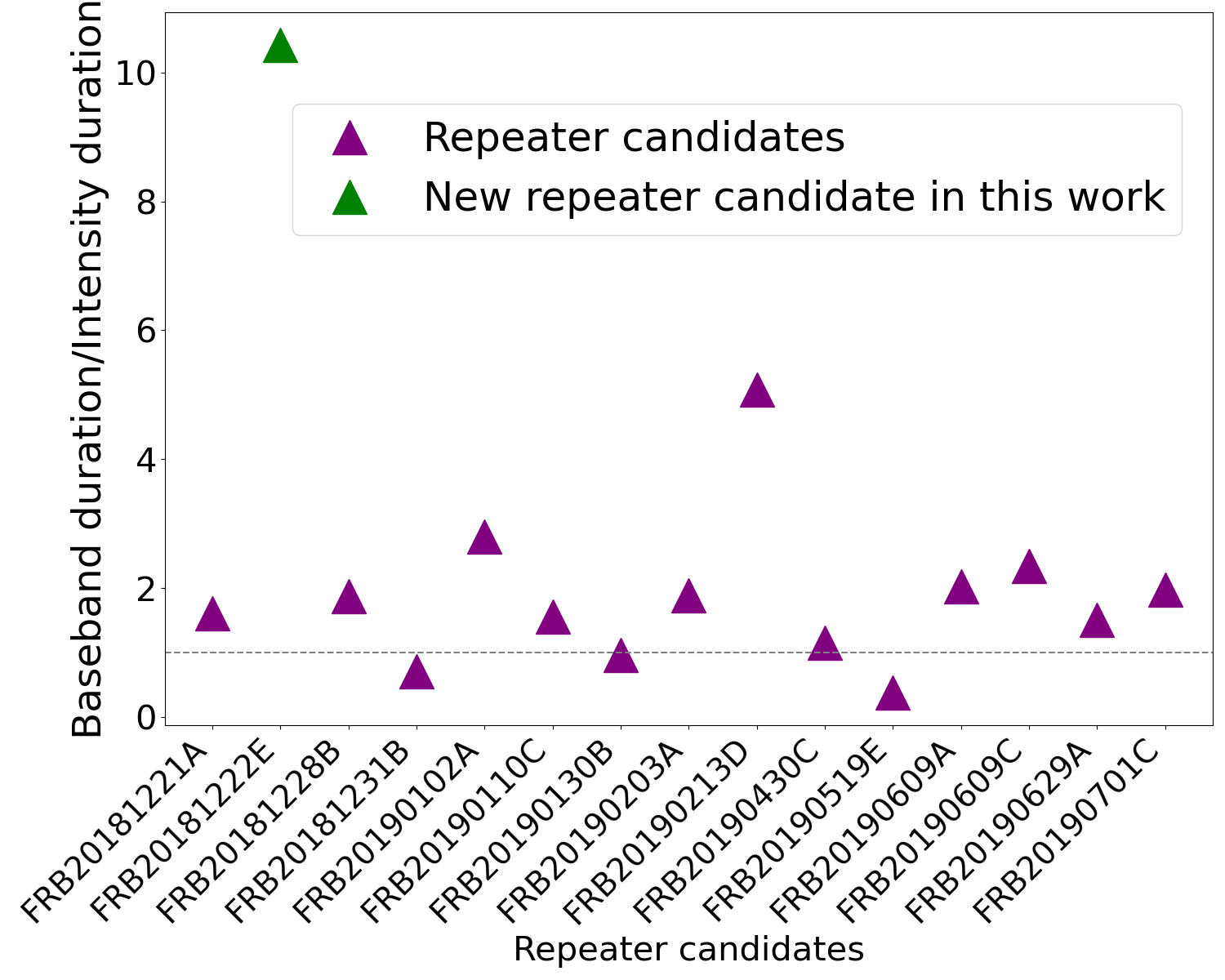}
\caption{
The ratios of the duration (s) in the baseband catalog to those in the intensity catalog for the repeater candidates identified in this work.
The red dot indicates the new repeater candidate identified in this work.
The blue dots show common repeater candidates between this work and \citet{2022MNRAS.509.1227C}.
}
\label{fig:duration_ratio}
\end{figure}

We contrasted the intensity duration with the baseband duration used in our study. 
For all 133 samples, we compared the durations in the baseband and intensity catalogs and found that 60.47\% have wider durations in the baseband catalog.
80\% of our repeater candidates have a wider duration in baseband than their intensity data. 
The 80\% is significantly higher than 60.47\% for the entire sample, indicating the importance of duration in identifying repeater candidates.
In detail, our repeater candidates are on average 2.85 times wider in baseband duration than intensity duration. 
Fig. \ref{fig:duration_ratio} provides a graphic representation of this contrast. 
Notably, FRB 20181222E, a new candidate that emerges as the widest among all candidates, with its baseband duration extending 10.4 times wider than its intensity duration. 
These results clearly show that our repeater candidates have a wider duration in baseband measurements. 

In contrast, three repeater candidates have shorter baseband durations than their intensity duration in Fig. \ref{fig:duration_ratio}. 
Figure \ref{fig:duration_scatter} presents a scatter plot with a 1:1 identity line to compare the durations from intensity and baseband data for repeaters and repeater candidates. 
Interestingly, we found that three confirmed repeaters also show shorter duration in baseband than in intensity data.
This characteristic of the confirmed repeaters may lead to our result that the three repeater candidates show shorter baseband duration than the intensity duration in Fig. \ref{fig:duration_ratio}. 
Overall, the wider baseband duration of our repeater candidates supports the conclusion in \cite{2021ApJ...923....1P}, which also enhances the reliability and coherence of our findings.

\FloatBarrier
\begin{figure}[tbp]
\centering
\includegraphics[width=0.9\linewidth]{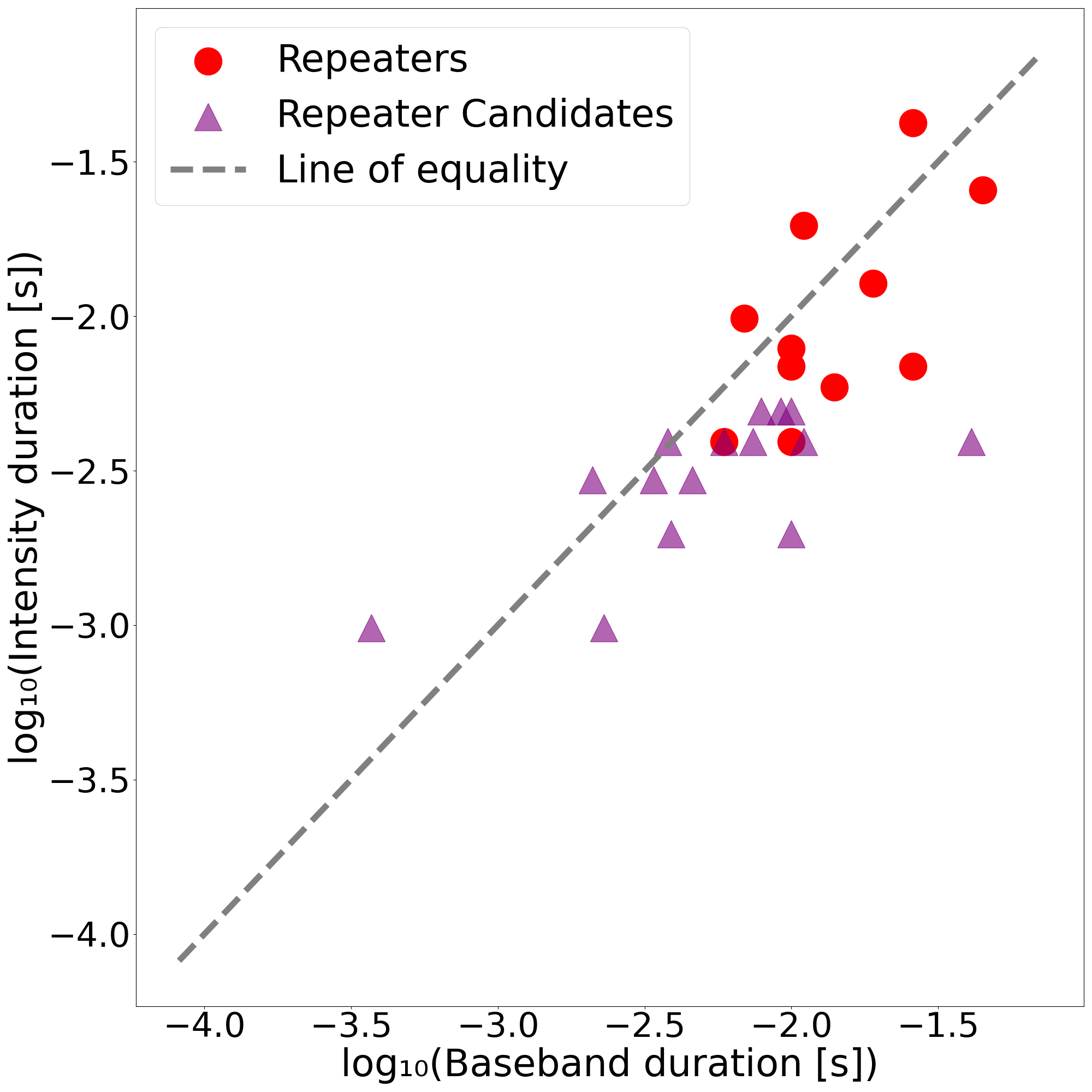}
\caption{
Comparison of baseband and intensity durations of repeater candidates and true repeaters.
The confirmed repeaters are shown by red dots, while repeater candidates are shown by blue dots.
The grey dashed line indicates the 1:1 relation.
}
\label{fig:duration_scatter}
\end{figure}

\subsubsection{31 conflict FRBs}\label{conflict}
One significant difference between our result and that of \cite{2022MNRAS.509.1227C}, as covered in section \ref{compare}, is the classification of the 31 conflict FRBs. 
In their study, these FRBs were found to be repeater candidates, whereas in ours, they were not identified as repeater candidates. 
In order to explore this discrepancy, we looked at the duration measurements of FRBs in baseband and intensity catalogues. 

As discussed in section \ref{duration}, in total samples, 60.47\% have wider durations in the baseband catalog. 
This means $100\%-60.47\% = 39.53\%$ of the total samples show shorter durations in the baseband catalog.
Additionally, a breakdown per category offers important insights.
In the conflict samples, 63.33\% of the 31 conflict samples exhibit shorter duration in the baseband catalog. 
The fraction of FRBs showing shorter duration in the baseband catalog is significantly higher in the 31 conflict samples (63.33\%) than that of the entire sample (39.53\%).   
This result implies that, in baseband measurements, the 31 conflict FRBs typically have a shorter duration than their intensity counterparts. 

\begin{figure}[tbp]
\centering
\includegraphics[width=1\linewidth]{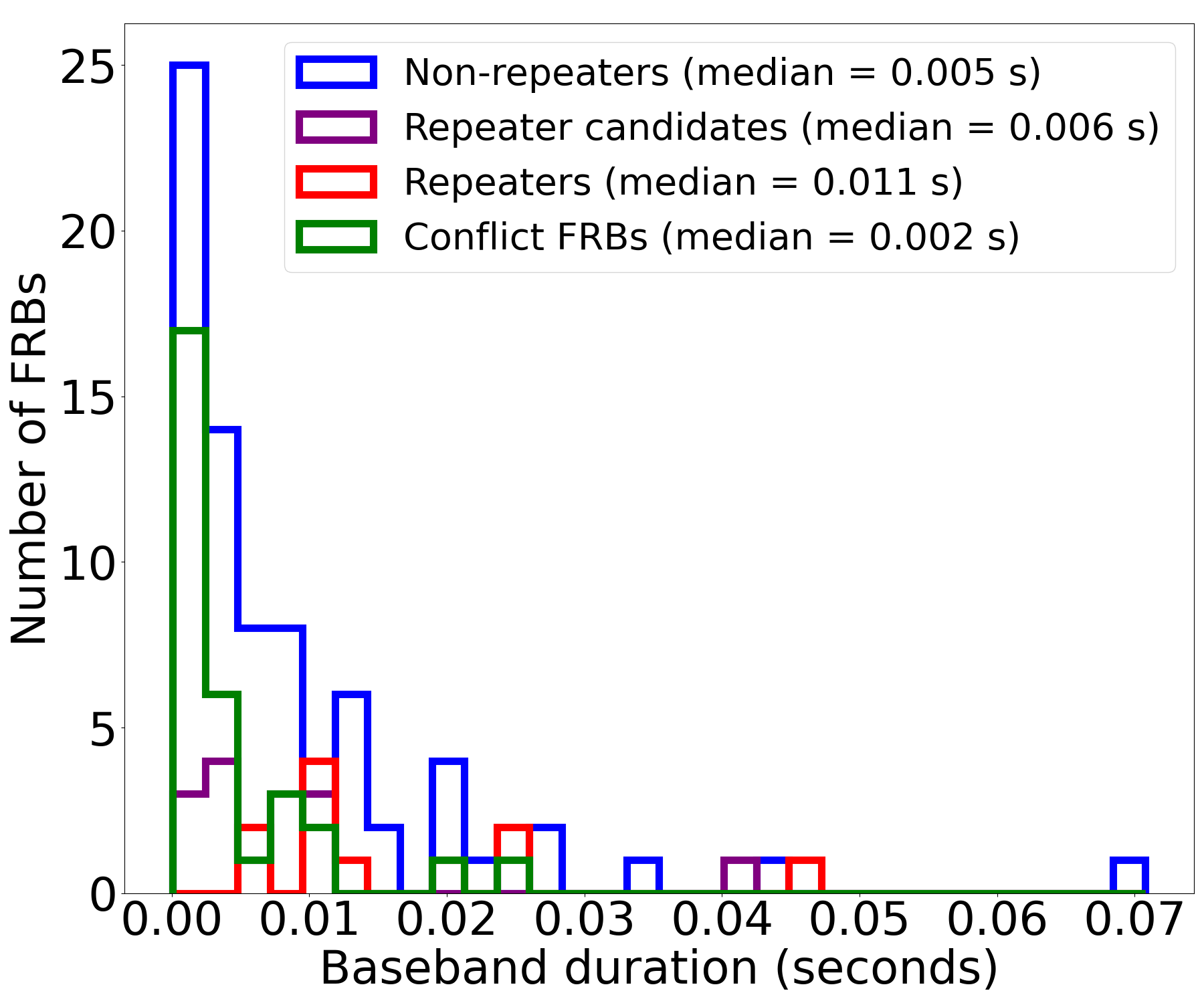}
\caption{
Histograms of durations of the 133 FRB samples in the baseband catalog.
Confirmed repeaters have the longest median in baseband duration (green) compared to all other groups. 
Among non-repeaters (blue), repeater candidates (orange), and conflict samples (red), the repeater candidates have a higher median duration than the others. 
This indicates that the duration of repeater candidates is similar to that of repeaters.
}
\label{fig:duration_histogram}
\end{figure}

Based on this change, our machine-learning model supports classification of the 31 FRBs as non-repeaters, offering compelling proof of our machine learning model capturing the wider duration FRBs for repeater candidates and the shorter duration FRBs for non-repeaters.  
We speculate \cite{2022MNRAS.509.1227C} could have misclassified the 31 FRBs as repeater candidates due to the wider duration in the intensity catalog that turned out to be narrower in baseband. 

\subsubsection{76 non-repeaters}\label{non-repeaters}
For the 76 common non-repeater samples, 61.84\% exhibit wider durations in the baseband catalog, and 34.21\% exhibit shorter durations in the baseband catalog. 
3.95\% exhibit exactly equal duration in both catalogs. 
The fraction of 61.84\% is similar to the value of the entire sample (60.47\%).
This is expected because the majority of our sample is non-repeaters in both \cite{2022MNRAS.509.1227C} and this work. Additionally, the distributions of baseband duration for non-repeaters, repeater candidates, repeaters, and conflict FRBs are shown in Fig. \ref{fig:duration_histogram}. Repeaters have the highest median value, followed by repeater candidates, non-repeaters, and conflict FRBs showing smaller medians.

\section{Conclusion}\label{conclusion}

Machine learning offers significant advantages in the study of FRBs. 
It can efficiently handle a large number of parameters and could facilitate the classification of FRBs without requiring long-term monitoring or extensive human intervention. 
Furthermore, if machine learning models are successful in detecting repeating FRB candidates, there is less need for extensive observational campaigns to verify that they repeat. In other words, observational efforts can be directed toward the specific bursts identified by machine learning models as potential repeating FRB candidates.


In this work, we performed an unsupervised machine learning classification of repeaters and non-repeaters with UMAP and HDBSCAN based on the CHIME baseband catalog. From our results, we found that the known repeaters form distinct clusters. We also identified some non-repeaters located within this cluster. These non-repeaters are considered repeater candidates, as they exhibit physical properties similar to those of known repeaters. Among our identified candidates, 14 overlap with those reported by \cite{2022MNRAS.509.1227C}, and we additionally discovered one new repeater candidate. However, 31 of the repeater candidates proposed by \cite{2022MNRAS.509.1227C} lie outside the repeater cluster in our analysis. This suggests that they do not share similar physical characteristics with known repeaters, and thus, we exclude them from the list of repeater candidates.

Compared with \cite{2022MNRAS.509.1227C}, our work offers several improvements. First, \cite{2022MNRAS.509.1227C} used the CHIME intensity catalog. On the other hand, we utilize the dataset with improved measurements obtained from the baseband catalog of the CHIME/FRB collaboration. Additionally, \cite{2022MNRAS.509.1227C} included several highly correlated features; we intentionally excluded such features to enhance the robustness and generalizability of our machine learning model. Furthermore, they selected the machine learning hyperparameters and repeater threshold for classification in an arbitrary manner. In contrast, our study systematically optimized both the hyperparameters and the threshold, improving model reliability.


Our repeater candidates show a wider duration in the baseband catalog than their intensity counterparts. In baseband measurements, these repeater candidates exhibit an average duration that is 2.85 times wider. Among the repeater candidates, our new candidate stands out because its baseband duration is 10.43 times wider than its intensity duration, making it the contender with the widest duration. 
Furthermore, the CHIME/FRB collaboration has confirmed one of our common repeater candidates as a repeater. 
The 31 FRBs excluded from the repeater candidates in this work show shorter durations than those in the intensity catalog.
In light of these arguments, we suggest conducting follow-up observations on these 14 repeater candidates (a new and remaining 13 repeating candidates) in order to verify the nature of their repetition.

\begin{acknowledgments}
\textbf{ACKNOWLEDGMENTS}\\
 We appreciate the referee's insightful comments, which improved the quality of the manuscript significantly.
TH is very grateful to the Ministry of Science and Technology of Taiwan through grants 113-2112-M-005-009-MY3, 113-2123-M-001-008-, 111-2112-M-005-018-MY3, and the Ministry of Education of Taiwan through a grant 113RD109. 
We thank National Chung Hsing University, Taiwan, for providing the required facilities to work on this project. W.J.P. has been supported by the Polish National Science Center project UMO-2020/37/B/ST9/00466. TG acknowledges the support of the National Science and Technology Council of Taiwan (NSTC) through grants 113-2112-M-007 -006, 113 -2927-I-007 -501, 113-2123-M-001 -008. This research was conducted under the agreement on joint mobility projects for the years 2024-2025 between the Polish Academy of Sciences and the National Science and Technology Council in Taiwan.
\end{acknowledgments}

%

\vspace{5mm}
\facilities{CHIME/FRB}


\software{UMAP \citep{2018arXiv180203426M}, HDBSCAN \citep{2019arXiv191102282M}, Scikit-learn \citep{JMLR:v25:19-301}, numpy \citep{2020Natur.585..357H}, pandas \citep{mckinney2010data}.
          }

\bibliography{PASPreference}{}
\bibliographystyle{aasjournal}



\end{document}